\documentclass[11pt,a4paper]{article}

\usepackage[a4paper,margin=1in]{geometry}
\usepackage{microtype}
\usepackage{times} 
\usepackage{amsmath,amssymb,bm}
\usepackage{graphicx}
\usepackage{booktabs,threeparttable}
\usepackage{indentfirst}
\usepackage{enumitem}
\usepackage{caption}
\usepackage{makecell}
\usepackage[numbers]{natbib}
\usepackage{array}
\usepackage{ragged2e}
\newcolumntype{P}[1]{>{\RaggedRight\arraybackslash}p{#1}}
\newcolumntype{C}[1]{>{\centering\arraybackslash}m{#1}}
\newcolumntype{L}[1]{>{\RaggedRight\arraybackslash}m{#1}}

\usepackage{multirow}

\usepackage{tcolorbox}
\tcbuselibrary{skins, breakable}
\usepackage{upgreek}
\usepackage{slashed}

\usepackage[colorlinks=true,
  linkcolor=blue!50!black,
  citecolor=blue!50!black,
  urlcolor=blue!50!black ,
   linktoc=page
]{hyperref}

\usepackage{titling}   
\pretitle{\begin{center}\LARGE\bfseries}
\posttitle{\par\end{center}\vspace{0.5em}}
\preauthor{\begin{center}}
\postauthor{\par\end{center}\vspace{0.5em}}

\usepackage{authblk}
                  
\usepackage[small,compact]{titlesec}
\titlespacing*{\section}{0pt}{0.9em}{0.4em}
\titlespacing*{\subsection}{0pt}{0.7em}{0.3em}
\titlespacing*{\subsubsection}{0pt}{0.5em}{0.2em}

\usepackage{abstract}

\usepackage{tocloft}
\usepackage{subcaption}

\usepackage{tikz}
\usetikzlibrary{arrows.meta,decorations.markings,decorations.pathmorphing,positioning,shapes.misc, matrix} 
\usepackage{tcolorbox}
\tcbuselibrary{skins, breakable}
\usepackage{tikz-feynman}
\tikzfeynmanset{compat=1.1.0}

\tikzset{
  fermion/.style={line width=0.8pt,
    postaction={decorate},
    decoration={markings, mark=at position 0.60 with {\arrow{Latex}}}},
  antifermion/.style={line width=0.8pt,
    postaction={decorate},
    decoration={markings, mark=at position 0.40 with {\arrow{Latex[reversed]}}}},
  diquark/.style={dashed, line width=0.8pt},
  overdraw/.style={preaction={draw,white,line width=2.0pt}}, 
  blob/.style={circle, draw, fill=gray!25, minimum size=9pt, inner sep=0pt}
}

\date{}
\title{\textbf{Experimental Tests of Baryon  and Lepton \\ Number Conservation}\\[0.5em]
\large Review for the \textit{Encyclopedia of Particle Physics}}
\vspace{3 em}
\author[1,2,3,4~$\dagger$]{~~~~~~Volodymyr Takhistov }
 
\affil[1]{International Center for Quantum-field Measurement Systems for Studies of the Universe  and Particles  \newline   (QUP, WPI), High Energy Accelerator Research Organization (KEK), Tsukuba, Japan}
\affil[2]{Theory Center, Institute of Particle and Nuclear Studies (IPNS), High Energy Accelerator \newline Research Organization (KEK), Tsukuba, Japan}
\affil[3]{Graduate University for Advanced Studies (SOKENDAI), Tsukuba, Japan}
\affil[4]{Kavli Institute for the Physics and Mathematics of the Universe (Kavli IPMU, WPI),   The University \newline  of Tokyo, Kashiwa, Japan} 
\begin{document}
 
\pretitle{\begin{center}\LARGE\bfseries}
\posttitle{\par\end{center}\vspace{3em}}

\setlength{\droptitle}{1.5em}
\maketitle  

\begingroup
\renewcommand\thefootnote{$\dagger$}\footnotetext{%
\small \href{mailto:vtakhist@post.kek.jp}{vtakhist@post.kek.jp}}
\endgroup

\thispagestyle{empty}

\renewcommand{\abstractname}{}   
\renewcommand{\absnamepos}{empty}

\vspace{-2em}
 
\begin{abstract}
\noindent
\textbf{Abstract:} Baryon number ($B$) conservation underlies the apparent stability of ordinary matter by forbidding the decay of nucleons, while lepton number ($L$) conservation plays a central role in the structure of lepton interactions and the possible origin of neutrino mass. In the Standard Model, $B$ and $L$ are accidental global symmetries rather than imposed fundamental principles. However, they are expected to be violated in many extensions of the theory, including frameworks of unification and processes in the early Universe. This review summarizes the status of experimental tests of $B$ and $L$ conservation and
discusses them within a unified framework for interpreting current and future searches across different processes and experimental approaches, outlining historical and theoretical motivation, key physical processes, as well as their broader connections and complementarity to other searches.
\end{abstract}

\vspace{5em}

\noindent
Keywords:
\textit{  
Accidental symmetries
- Baryon number violation
- Cosmological baryon asymmetry
- Effective field theory
- Grand unification
- Lepton number violation
- Neutrinoless double beta decay
- Neutrino mass
- Neutron-antineutron oscillations
- Nucleon decay
- Physics beyond the Standard Model
- Stability of matter
}

\clearpage

\begingroup
\hypersetup{linkcolor=blue!50!black}
\tableofcontents
\endgroup

\vspace{1em}

\begin{tcolorbox}[
  breakable,
  colback=gray!10,
  colframe=gray!60,
  title=\textbf{Conceptual Insights},
  fonttitle=\bfseries,
  left=1.2em,
  right=1.2em,
  top=0.8em,
  bottom=0.8em
]
\begin{itemize}[
  label=\raisebox{0.2ex}{\large$\bullet$},
  leftmargin=1.2em,
  labelsep=0.6em
]

\item Baryon and lepton number conservation in the Standard Model of particle physics emerge as accidental structures rather than as fundamental principles.~Their violation indicates qualitatively new physics.

\item Baryon and lepton number violation probe organizing principles of the underlying fundamental theory, including those associated with unification and the origin of the cosmological matter-antimatter asymmetry, in regimes often challenging to access experimentally.

\item Observation of baryon number violation establishes that baryon number is not an exact symmetry protecting the stability of ordinary (baryonic) matter.
Observation of lepton number violation by two units in processes such as neutrinoless double beta decay would provide compelling evidence that neutrinos and antineutrinos are not distinct particles.

\item The non-observation of such symmetry violations at increasingly high precision sets powerful constraints on possible extensions of known physics, disfavoring simple mechanisms and guiding theoretical developments toward more subtle, complex or higher energy origins.

\item Different physical processes and channels of potential symmetry violation enable distinct tests of underlying organizing structures, leading to a natural complementarity among observational approaches and technologies.
  
\end{itemize}
\end{tcolorbox}

\section{Introduction } 

The remarkable stability of ordinary baryonic matter is a striking empirical fact of Nature.
Atoms, nuclei and macroscopic matter can persist on timescales vastly exceeding the age of the Universe,
enabling chemistry, astrophysical structure formation and the emergence of life.
At the microscopic level, this stability reflects both  dynamical mechanisms associated with strong force interactions and nuclear binding, as well as selection rules associated with
approximate  conservation of quantum numbers that forbid otherwise kinematically allowed decay processes. Experimental searches for violations of these conservation laws  have long served as uniquely sensitive probes of physics beyond the Standard Model (SM) of particle physics, capable of accessing effective energy scales far beyond the typical direct reach of experimental techniques such as particle colliders.

The modern conceptual understanding of baryon number ($B$) and lepton number ($L$) conservation has emerged gradually. 
These quantum numbers, defined respectively as the number of baryons (leptons) minus antibaryons (antileptons), summarize  powerful empirical facts and observed patterns. 
While Rutherford's 1911 scattering experiments established the existence of compact positively charged atomic nuclei~\cite{Rutherford:1911zz}, subsequent developments in quantum mechanics and nuclear physics clarified how such nuclei can be bound and not disintegrate despite the electromagnetic repulsion implied by their concentrated positive charge.
However, the stability of the lightest baryon, the proton, posed a deep question.

Without an additional forbidding principle,  proton decay into lighter states consistent with Lorentz
invariance and electric charge conservation  could in principle be allowed.
To account for the empirical stability of protons and matter, the conservation of $B$ was introduced as a
phenomenological selection rule. $B$ conservation ensures the stability of the lightest baryon, the proton, while allowing ordinary weak decays of heavier baryons, such as neutron beta decay, which conserve $B$. A form of such symmetry was considered by Weyl in 1929~\cite{Weyl:1929}, and the concept was further developed by 
Stueckelberg~\cite{Stueckelberg:1938zz} and Wigner~\cite{Wigner:1949}.
In a parallel development, an additive $L$ was introduced to encode its apparent
conservation in weak interactions and the distinct identities of charged leptons
and neutrinos. A central question is whether these organizing principles reflect the fundamental symmetries of Nature or are merely approximate emergent consequences at low energies.

In SM, $B$ and total $L$ are conserved to very high accuracy in laboratory processes. This apparent conservation is not imposed by a fundamental principle such as gauge symmetry. Instead, $B$ and $L$ emerge as accidental global symmetries arising from SM's field content and interaction structure.
As a result, violations of $B$ and $L$ are not fundamentally forbidden and may arise from physics beyond SM, including through new interactions or organizing structures.
These violations can manifest in a variety of processes such as proton decay, neutron-antineutron  $(n-\overline{n})$ transitions and $L$-violating phenomena like neutrinoless double beta decay $(0\nu\beta\beta)$.
Such processes can be systematically classified according to the changes they impose on $B$ and $L$, allowing for a unifying framework for interpretations and experiments.

The history of fundamental physics 
provides multiple precedents in which apparent symmetries inferred from low energy observations and once regarded as exact later found to be violated. Parity ($P$), the symmetry related to spatial inversions, was shown to be violated in weak interactions
through the beta decay experiments of Wu et al. in 1957~\cite{Wu:1957my}, following the proposal by Lee and Yang in 1956~\cite{Lee:1956qn}.
This was followed by the realization that charge conjugation ($C$) is also violated.
Prominently, the combined $CP$ symmetry was found to be violated in neutral kaon decays by Christenson, Cronin, Fitch and Turlay in 1964~\cite{Christenson:1964fg}.
These discoveries fundamentally reshaped our understanding of symmetries and their role. They established
that conservation laws appearing exact at accessible energies may fail in other regimes, 
revealing new principles.

Notably, even within the SM itself $B$ and $L$ are not exact symmetries.
Electroweak chiral anomalies~\cite{Adler:1969gk,Bell:1969ts} imply non-conservation of $B+L$ through non-perturbative
$SU(2)_L$ gauge field  configurations, while  preserving the combination $B-L$ within the SM~\cite{tHooft:1976up}.
At zero temperature such processes are exponentially suppressed and are thus not typically relevant for
laboratory searches. However, at temperatures above the electroweak scale they become unsuppressed and
can efficiently violate $B$ and $L$ in equal amounts while preserving $B-L$~\cite{Kuzmin:1985mm}.
This highlights connections between laboratory tests of $B$ and $L$ conservation and cosmology. In particular, $B$ violation constitutes
one of the key conditions identified by Sakharov~\cite{Sakharov:1967dj} required to generate the observed matter-antimatter asymmetry of the Universe, as inferred from precise cosmological observations.

New physics beyond the SM associated with leptons is independently motivated by the discovery of neutrino oscillations by the Super-Kamiokande experiment in 1998~\cite{Super-Kamiokande:1998kpq}, which implied that neutrinos possess non-zero masses. If neutrinos are their own antiparticles and are Majorana fermions, the presence of neutrino masses necessarily implies 
$L$ violation by two units with $\Delta L = 2$. In a variety of theoretical frameworks  the interactions responsible for generating Majorana neutrino masses could also provide the ingredients required for leptogenesis mechanisms capable of generating the cosmic baryon asymmetry~\cite{Fukugita:1986hr}.
Conversely, if neutrinos are purely Dirac particles $L$ may be conserved as an exact global symmetry or as a remnant of a gauged symmetry, such as gauged $B-L$. However, a variety of theories extending the SM predict experimentally accessible
$\Delta L \neq 0$ processes. Notably, while neutrino oscillations conserve total $L$  these processes provide direct evidence for the violation of individual lepton flavor numbers $L_e$, $L_{\mu}$, $L_{\tau}$ associated with the corresponding charged leptons. Therefore, experimental tests of $L$ conservation are intimately linked to the origin of neutrino masses and probe the structure of underlying physical principles at different energy scales and regimes.

From a ``top-down'' viewpoint of fundamental theory, violations of quantum numbers associated with global symmetries can be expected to be generic.
For example, Grand Unified Theories (GUTs) that unify the SM gauge interactions and embed quarks and leptons within a single framework predict in many realizations
$\Delta B = 1$ nucleon decay processes such as $p \to e^+ \pi^0$ associated with effective unification scales
$\Lambda \sim 10^{15-16}~ \mathrm{GeV}$~\cite{Georgi:1974sy,Fritzsch:1974nn}.
Expectations for $B$ and $L$ violation also arise in a broad range of other SM extensions  including theories with additional spacetime dimensions, extended gauge sectors and supersymmetry (SUSY). Further, a variety of arguments suggest that a consistent theory of quantum gravity does not admit exact global
symmetries, implying that global $B$ and $L$ symmetries should be violated at some level unless
protected by gauge symmetries~\cite{Banks:2010zn,Harlow:2018tng}.
The combined $B-L$ symmetry plays  a special role as it is anomaly free in SM when right handed neutrinos are included and can thus be
gauged, providing a minimal route to suppress or correlate $B$-violating and
$L$-violating effects~\cite{Mohapatra:1980qe}.

These considerations motivate a broad experimental program targeting processes that violate $B$ or $L$, including $\Delta B = 1$, $\Delta B = 2$ and $\Delta L = 2$ transitions, across a wide range of detector technologies and experimental techniques. Beyond conventional channels, $B$ and $L$ violation could also manifest through processes involving new invisible or weakly interacting final states, as well as through processes induced or catalyzed by external interactions.
Experimentally, rare process searches for $B$ and $L$ violation enable probing extraordinarily high effective energy scales by exploiting large detector target masses, long observation exposures and very low backgrounds. Historically, experiments originally developed to test the stability of matter have simultaneously enabled transformative discoveries in other frontiers, such as neutrino physics. This highlights the deep interplay among $B$ and $L$ conservation tests. Notably, these searches are also closely related to other tests of physics beyond SM, such as probes of charged lepton flavor violation (CLFV) that conserve total $L$ but provide complementary sensitivity to the structure of new physics in the lepton sector.

\section{Motivation and Theoretical Framework}

\subsection{Accidental symmetries of the Standard Model}
\label{subsec:accidental_symmetries}

In the SM with gauge group $SU(3)_C \times SU(2)_L \times U(1)_Y$, $B$ and $L$ are global quantum numbers assigned to SM matter (fermion) quantum fields 
as 
\begin{equation}\begin{cases}
\textrm{Baryon number:}~
B =~{\rm 1/3} ~~~ \text{for quarks }~q~,    \\[4pt]
\textrm{~Lepton number:}~L =~ 1 ~~~ \text{for leptons }~l~.
\end{cases}
\end{equation} 
Here, $q$ denotes the up $u$, down $d$, charm $c$, strange $s$, top $t$  and bottom $b$
quark fields, while $l$ denotes the charged leptons electron $e^-$, muon $\mu^-$, tau $\tau^-$ and their corresponding
neutrino fields $\nu_e$, $\nu_\mu$ and $\nu_\tau$.
Antiquark fields $\bar q$ carry $B=-1/3$ and antilepton fields $\bar l$ carry $L=-1$.
All SM gauge bosons and the Higgs field carry $B=L=0$.

$B$ and $L$ quantum numbers are not imposed by any fundamental symmetry or principle in SM.
Instead, $B$ and $L$ emerge as accidental global symmetries  of the renormalizable Lagrangian~\cite{Weinberg:1995mt}. In particular, given the SM field content and gauge invariance there are no Lorentz  and
gauge invariant operators of dimension $d \le 4$ that violate either $B$ or $L$.
All renormalizable interactions are found to automatically conserve these quantum numbers. This includes the Yukawa couplings responsible for SM fermion masses.
Thus, proton stability  in SM is an emergent consequence.

\begin{figure}[t]
\begin{subfigure}[b]{0.32\linewidth}
  \centering
  \makebox[\linewidth][c]{\includegraphics[height=2.9cm]{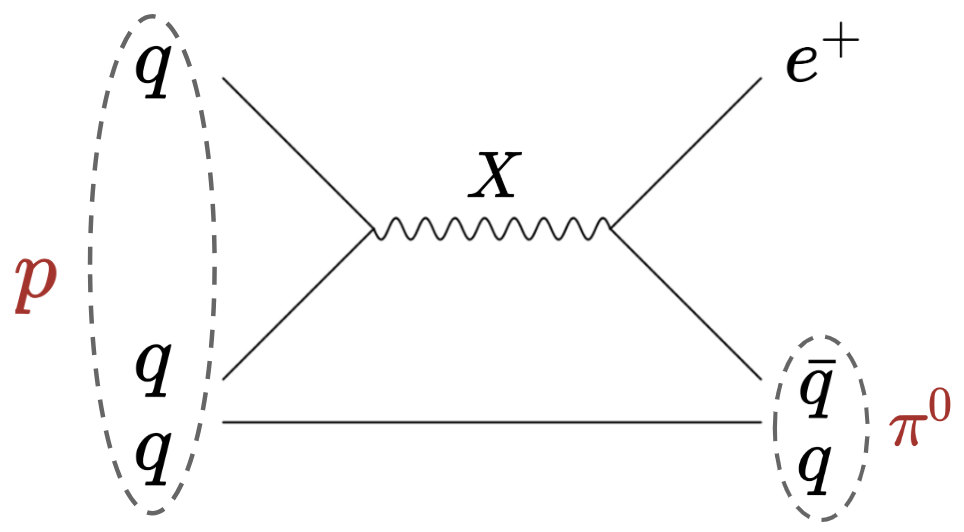}} 
  \caption{$p \to e^+\pi^0$\\\hspace{2mm}($\Delta B=1,\ \Delta L=1$)}
\end{subfigure} 
\hfill
 \begin{subfigure}[b]{0.28\linewidth}
  \centering
    \makebox[\linewidth][c]
  {\includegraphics[height=1.0cm]{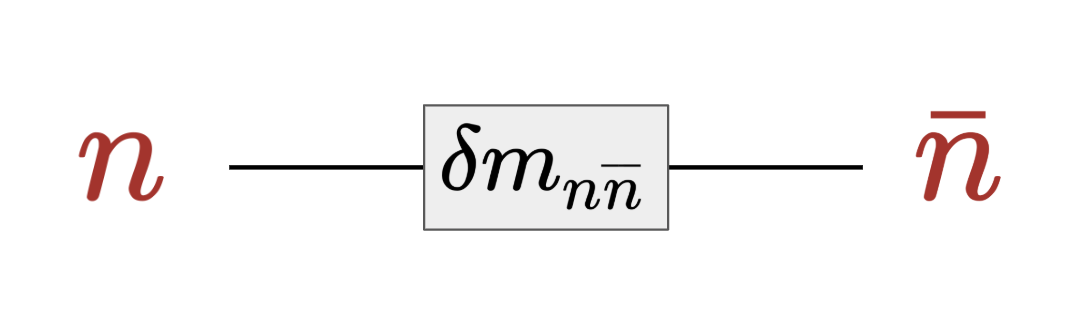}}  \vspace{1.5em}
  \caption{$n - \bar n$ \\ \hspace{3mm} ($\Delta B=2$)}
\end{subfigure}
\hfill
\begin{subfigure}[b]{0.32\linewidth}
  \centering
  \makebox[\linewidth][c]
  {\includegraphics[height=4.2cm]
  {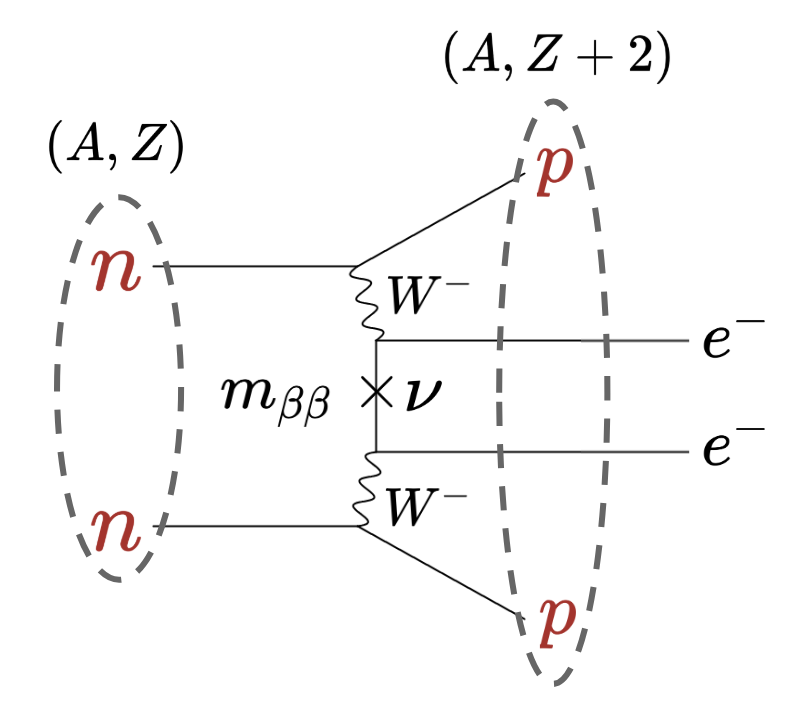}}\vspace{-0.8em}
  \caption{$0\nu\beta\beta$ \\\hspace{3mm}($\Delta L=2$)}
\end{subfigure}
\caption{Representative processes of baryon number ($B$) and lepton number ($L$) violation.
\textbf{(a)} Proton decay illustrated by $p \to e^{+}\pi^{0}$ that violates $\Delta B = 1$ and $\Delta L = 1$.
\textbf{(b)} Neutron-antineutron ($n-\bar n$) oscillations that induce $\Delta B = 2$ violation. 
\textbf{(c)} Neutrinoless double-beta decay ($0\nu\beta\beta$) that induces $\Delta L = 2$  in nuclear transitions.}
\label{fig:processes}
\end{figure}

At the quantum level, $B$ and $L$ are not exactly conserved in SM.
While the classical currents associated with $B$ and $L$ are conserved, both symmetries are violated
by chiral anomalies involving the electroweak $SU(2)_L$ gauge fields~\cite{tHooft:1976up} (see also  Ref.~\cite{Klinkhamer:1984di} for sphalerons).
Non-perturbative instanton and sphaleron gauge field configurations provide the realization
of this anomaly and induce  transitions with $\Delta(B+L) \neq 0$ while preserving $\Delta(B-L)=0$,
leading to $B$ and $L$ violation by equal amounts with $\Delta B = \Delta L = 3 \Delta N_{\rm CS}$ where $\Delta N_{\rm CS} $ is the integer-valued change in Chern-Simons number of the $SU(2)_L$ gauge fields for the three fermion generations of SM.
Although these effects are exponentially suppressed at zero temperature, they can become efficient
in the early Universe at temperatures above the electroweak symmetry breaking scale where they may play a prominent role in
scenarios of baryogenesis~\cite{Rubakov:1996vz}.

As a global symmetry, the combination $B-L$ is free from anomalies associated with SM gauge interactions. Many SM extensions consider promoting $B-L$ to a local $U(1)_{B-L}$ gauge symmetry, which for consistency typically requires additional fields (e.g. right handed neutrinos) to cancel the resulting gauge anomalies. 
Notably, beyond gauging the anomaly free combination $B-L$  it is also possible to promote $B$ and $L$ individually to local gauge symmetries $U(1)_B$ and $U(1)_L$ in SM extensions by introducing additional fields that cancel the associated gauge anomalies~\cite{FileviezPerez:2010gw}.
Once $B-L$ is violated (e.g. by two units), processes with $\Delta L = 2$ can arise, enabling Majorana neutrino masses with neutrinos as their own antiparticles and also processes such as neutrinoless
double beta decay.
Violation of $B-L$ also permits qualitatively distinct $B$-violating processes, including
$\Delta B = 2$ transitions such as $n- \bar{n}$ oscillations that probe different symmetry
structures and energy scales from those typically associated with   proton decay.

These considerations highlight that $B$ and $L$ conservation in the SM is neither fundamental nor exact.
Instead, $B$ and $L$ emerge as low energy features of the SM structure and are not protected against violation~\cite{Weinberg:1995mt}.
Such violations can arise through a variety of effects including non-perturbative dynamics, non-renormalizable operators of higher dimension $d > 4$ or more fundamental organizing principles at higher energies in which exact global symmetries are not conserved, as suggested by arguments from quantum gravity~\cite{Banks:2010zn,Harlow:2018tng}. 

This perspective motivates a broad experimental program searching for violations of $B$ and $L$ as sensitive probes of physics beyond the SM. 
Fig.~\ref{fig:processes} displays representative physical processes of $B$ and $L$ violation including proton decay $p \to e^{+}\pi^{0}$ ($\Delta B = 1$, $\Delta L = 1$), $n-\bar n$ oscillations ($\Delta B = 2$) and $0\nu\beta\beta$ decay in nuclear $(A,Z)\to(A,Z+2)$  transitions ($\Delta L = 2$), where $A$ and $Z$ denote the nuclear mass and atomic numbers, respectively. In extensions of the SM based on unification  proton decay  is typically mediated by very heavy states (e.g. GUT $X$ gauge bosons) and can be described at low energies by dimension $d=6$ effective operators. On the other hand, $n-\overline{n}$ oscillations can arise from higher dimensional quark  level effective operators and can be parameterized at the nucleon level by an effective mixing between the neutron $n$ and antineutron $\bar n$, characterized by an off-diagonal $n - \bar{n}$ mass mixing parameter $\delta m_{n\bar n}$. $L$ violation can be probed by $0\nu\beta\beta$ decay and  in many scenarios  its rate can be expressed in terms of the effective Majorana neutrino mass $m_{\beta\beta}$.
Together, these processes highlight the complementary experimental routes through which violations of $B$ and $L$ enable tests of fundamental physics beyond the SM across a broad range of energy scales and physical regimes.

\subsection{Effective field theory classification}

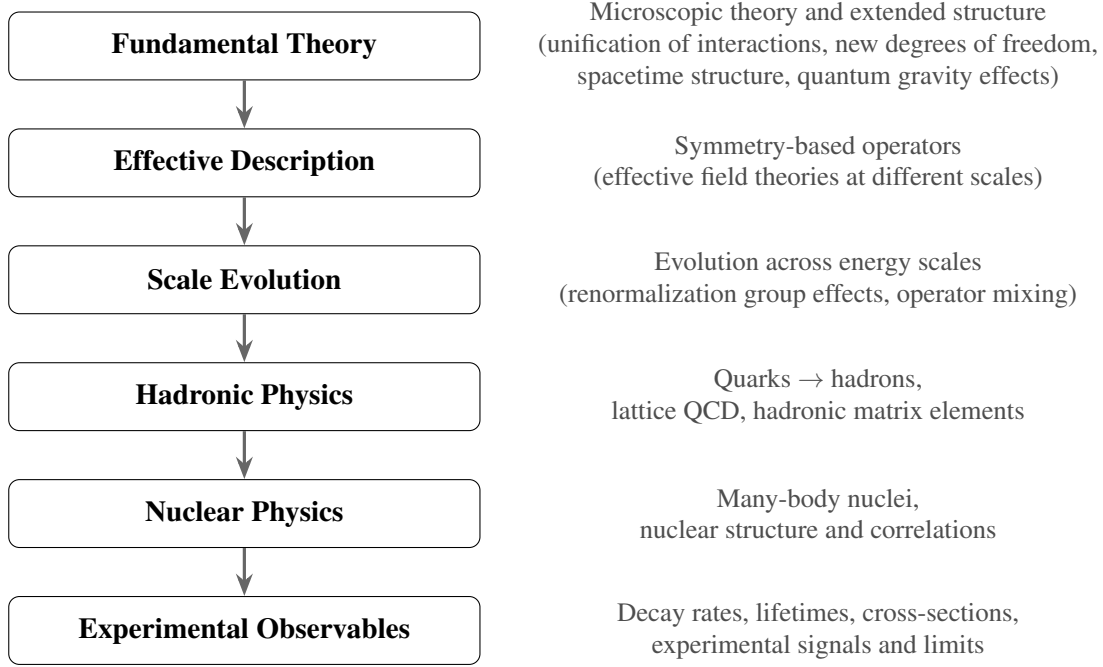
\begin{figure}[t]
\centering
\begin{tikzpicture}[
    node distance=1.55cm,
    every node/.style={draw, rounded corners, align=center},
    title/.style={font=\bfseries, minimum height=0.9cm, minimum width=6.2cm},
  desc/.style={draw = none,  font=\small\color{black!70},    text=black!70, minimum height=1.05cm, minimum width=6.2cm}, 
    arrow/.style={->, line width=1.2pt, >=Stealth, color=black!60}
] 
\node[title ] (UVt) {Fundamental Theory};
\node[title, below of=UVt] (Efft) {Effective Description};
\node[title, below of=Efft] (RGt) {Scale Evolution};
\node[title, below of=RGt] (Ht) {Hadronic Physics};
\node[title, below of=Ht] (Nt) {Nuclear Physics};
\node[title, below of=Nt] (Ot) {Experimental Observables};
\node[ desc, right of=UVt, xshift=6cm]
{Microscopic theory and extended structure \\ (unification of interactions, new degrees of freedom, \\spacetime structure, quantum gravity effects)};
\node[ desc, right of=Efft, xshift=6cm]
{Symmetry-based operators\\ (effective field theories at different scales)};
\node[ desc, right of=RGt, xshift=6cm]
{Evolution across energy scales \\ (renormalization group effects, operator mixing)};
\node[ desc, right of=Ht, xshift=6cm]
{Quarks $\rightarrow$ hadrons,\\ lattice QCD, hadronic matrix elements};
\node[ desc, right of=Nt, xshift=6cm]
{Many-body nuclei,\\ nuclear structure and correlations};
\node[ desc, right of=Ot, xshift=6cm]
{Decay rates, lifetimes, cross-sections,\\ experimental signals and limits};
\draw[arrow] (UVt) -- (Efft);
\draw[arrow] (Efft) -- (RGt);
\draw[arrow] (RGt) -- (Ht);
\draw[arrow] (Ht) -- (Nt);
\draw[arrow] (Nt) -- (Ot);
\end{tikzpicture}
\caption{Schematic flow from fundamental theory to experimental observables. Symmetry constraints and selection rules govern the allowed structures at each analysis stage, while effective descriptions provide a universal bridge between microscopic theory and experimentally accessible observables.}
\label{fig:theorytoexp}
\end{figure}

A unifying and systematic framework for relating microscopic physics of fundamental theories
at high  energies to the experimentally accessible observables at low  energies is provided by effective field theory (EFT). This connection
proceeds through a sequence of methodological steps. They involve the construction of effective
operators restricted by underlying symmetries of the theory, their renormalization group evolution across energy scales  and their matching onto hadronic
and nuclear degrees of freedom. Fig.~\ref{fig:theorytoexp} summarizes this framework spanning multiple energy scales and physical regimes, highlighting where theoretical inputs
including lattice quantum chromodynamics (QCD) and nuclear structure calculations enter the interpretation of experimental results.

From the EFT perspective the accidental perturbative conservation of $B$ and $L$
by the renormalizable SM Lagrangian implies that violating effects can first appear through
higher dimensional $d > 4$ operators suppressed by powers of a high energy scale $\Lambda$ where new physics beyond the SM has been integrated out~\cite{Weinberg:1979sa}.
The connection between microscopic dynamics and structures at high energies and experimentally
accessible observables at low energies can then be established through a sequence of
effective descriptions. This includes construction of gauge-invariant operators at energy scales
$\sim\Lambda$, renormalization group evolution to the electroweak scale including
possible operator mixing~\cite{Alonso:2013hga} and matching onto EFTs below the electroweak scale and ultimately onto hadronic and nuclear degrees of freedom.
For $B$ violation this procedure typically involves matching quark level operators
onto hadronic operators and,  where relevant,  embedding them into nuclear matrix elements, as in the case of nucleon decay.
For $L$ violation,  considering neutrinoless $0\nu\beta\beta$
as a representative example, the process may proceed either through the exchange of light
Majorana neutrinos over nuclear distances or through short distance contact interactions
induced by new physics at higher energy scales.
The latter, as well as $\Delta B=2$ processes such as $n-\bar{n}$ oscillations in nuclei,
require consistent matching onto chiral EFT operators and nuclear response functions
to describe their effects in nuclear environments~\cite{Cirigliano:2022oqy}.
 
\begin{figure}[t]
\centering
\includegraphics[width=1\linewidth,trim=1cm 1cm 1cm 1cm,
  clip]{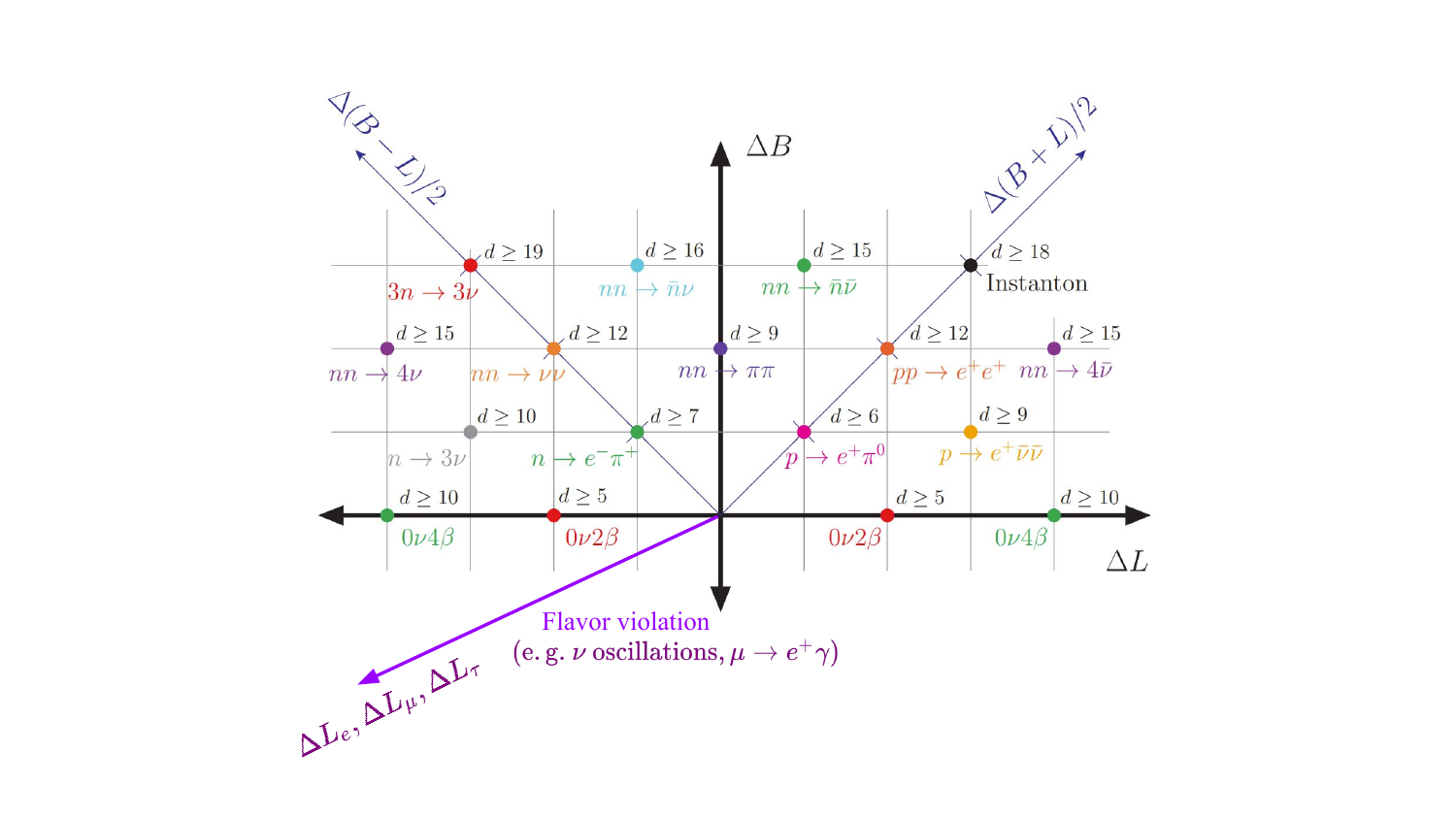}
\caption{Landscape of $B$-violating and $L$-violating transitions in the $(\Delta B,\Delta L)$ plane,
indicating the lowest SMEFT operator dimension $d$ at which each class can arise.
Adapted from Ref.~\cite{Heeck:2019kgr} and extended to incorporate
lepton flavor violation.}
\label{fig:landscape}
\end{figure}

At energies significantly below the scale $\Lambda$, $B$ and $L$ violation can be described systematically within
the framework of SM EFT (SMEFT) by augmenting the renormalizable local SM Lagrangian $\mathcal{L}_{\rm SM}$ with
higher dimensional gauge-invariant operators constructed from SM fields~\cite{Buchmuller:1985jz,Grzadkowski:2010es}
\begin{align}
\mathcal{L}_{\rm eff}
 = 
\mathcal{L}_{\rm SM}
 + 
\sum_{d>4}\sum_i \frac{C^{(d)}_i}{\Lambda^{d-4}} 
\mathcal{O}^{(d)}_i
 + {\rm h.c.},
\label{eq:SMEFTexp}
\end{align}
where $\mathcal{O}^{(d)}_i$ are gauge invariant operators of mass dimension $d$
forming a complete basis at each order in $1/\Lambda$, ``h.c.'' denotes Hermitian conjugation,
$C^{(d)}_i$ are dimensionless Wilson coefficients encoding short distance microscopic dynamics
and each operator can be related to a specific $(\Delta B,\Delta L)$ set.
Notably, this expansion is independent of the specific theoretical model. The imposed symmetry constraints related to the operator basis are separated from microscopic dynamics
encoded by Wilson coefficients  and thus the framework provides a unified description for comparing distinct processes.

Throughout, operators are written schematically to illustrate their quantum number structure and mass dimension. Color, Lorentz, gauge and flavor generation index contractions are implicit. Explicit operator bases, including full index contractions and operator redundancies, can be found for example in Refs.~\cite{Grzadkowski:2010es,Alonso:2013hga}.  The notation employed reflects the effective description discussed. SMEFT operators are shown schematically using left handed Weyl fields, with right handed fermions written as charge conjugated singlets. When discussing UV completions or phenomenological interactions, expressions are occasionally given in conventional four-component Dirac notation, while SUSY extensions are formulated in terms of superfields. This presentation is conventional and  does not affect the underlying physics.  Here, $L$ and $Q$ denote the left handed lepton and quark $SU(2)_L$ doublets, respectively, while $e^c$, $u^c$, and $d^c$ denote the charge-conjugated right handed charged lepton and up-type and down-type quark singlets, written as left handed Weyl fields. The field $H$ denotes the SM Higgs doublet  with $\tilde{H} = i\sigma_2 H^\ast$.

The possible $\Delta B$ and $\Delta L$ operators can be organized from consideration of Lorentz invariance and SM
gauge symmetry structure, 
given the SM field content and locality as well as optionally including right handed gauge-singlet neutrinos.
Any $SU(3)_C \times SU(2)_L \times U(1)_Y$ gauge-invariant SMEFT operator that violates $B$ or $L$ at dimension $d$ obeys relation~\cite{Kobach:2016ami}
\begin{align}
\frac{(\Delta B-\Delta L)}{2}\in \mathbb{Z} ,
\qquad
\begin{cases}
d~{\rm even}: & |\Delta B-\Delta L|=0,4,8,\ldots\\[2pt]
d~{\rm odd}:  & |\Delta B-\Delta L|=2,6,10,\ldots
\end{cases}
\label{eq:selection}
\end{align}
where $\mathbb{Z}$ denotes the set of integers. This holds
independently of the specific high energy (ultraviolet, UV) completion under the SMEFT assumptions.
Selection rules of Eq.~\eqref{eq:selection} can explain several generic features considered here:
(i) the lowest dimension $\Delta L=2$ interaction can appear at $d=5$,
(ii) the leading $\Delta B=1$ operators can arise at $d=6$ and typically conserve $B-L$, 
(iii) $\Delta B=2$ processes can first appear at higher $d = 9$ dimension   and probe distinct symmetry structures.

From these considerations it is apparent that transparent and experimentally useful organization of relevant physical processes can be obtained by their $(\Delta B,\Delta L)$ structure. Fig.~\ref{fig:landscape}, adapted from Ref.~\cite{Heeck:2019kgr} to incorporate lepton flavor violation, summarizes $B$-violating and $L$-violating processes in the
$(\Delta B,\Delta L)$ plane  indicating the lowest operator dimension at which each class of
transitions can arise and illustrating the role of selection rules.
This $(\Delta B,\Delta L)$ taxonomy provides a unifying framework for both theoretical interpretations and
experimental strategies.

The lowest dimensional representative operators in each class illustrate the hierarchy of effective operators, highlighting how $B$ and $L$ violations arise at different mass dimensions in SMEFT.
The unique dimension $d = 5$ operator constructed from SM fields violating $L$ by two units, with $\Delta L=2$, $\Delta B=0$ and $|\Delta(B-L)|=2$, is the Weinberg operator~\cite{Weinberg:1979sa}
\begin{align}
\mathcal{O}_5 \sim \frac{1}{\Lambda} (L \tilde{H})(L \tilde{H})~,
\label{eq:weinbergop}
\end{align}
which provides the leading effective interaction responsible for Majorana neutrino masses within the minimal SMEFT field content.
After electroweak symmetry breaking  when Higgs field acquires a vacuum expectation value $v \simeq 246$~GeV  this operator generates Majorana neutrino masses $m_\nu \sim v^2 / \Lambda$, implying that neutrinos are their own antiparticles. This enables $L$-violating transitions such as  $0\nu\beta\beta$ decay with violation $\Delta L = 2$, as illustrated in Fig.~\ref{fig:processes}. The Schechter-Valle theorem ensures that the observation of $0\nu\beta\beta$ implies the existence of a Majorana mass term for neutrinos within a local, Lorentz-invariant gauge theory, irrespective of the underlying microscopic mechanism~\cite{Schechter:1981bd}, although the induced mass can be highly suppressed and negligibly small compared to observed neutrino masses~\cite{Duerr:2011zd}.
At the level of UV completion, this operator can arise from a variety of mechanisms. This includes type-I seesaw with right handed neutrinos~\cite{Minkowski:1977sc,Yanagida:1979as,Gell-Mann:1979vob}, type-II seesaw involving heavy scalar Higgs triplets~\cite{Mohapatra:1980yp}, type-III seesaw with heavy Majorana fermion triplets~\cite{Foot:1988aq,Ma:1998dn}, as well as other extensions containing heavy degrees of freedom. Such models can provide intuition for the origin of $L$ violation and neutrino mass. However, the interpretation of experimental searches emphasized here is particularly transparent when organized in a model-independent EFT framework.

 \begin{table}[t]
\centering
\small
\renewcommand{\arraystretch}{1.25}
\setlength{\tabcolsep}{4pt}
\begin{tabular}{c|ccc|c|p{5cm}}
\Xhline{1.2pt}
\textbf{Operator  
} 
& \textbf{Dim. ($d$)} & $\Delta B$ & $\Delta L$ &
\textbf{Representative process} & \multicolumn{1}{c}{\textbf{Interpretation}} \\
\midrule
$\dfrac{(L \tilde{H})(L \tilde{H})}{\Lambda}$ & 5 & 0 & 2 &
$0\nu\beta\beta$ &
Weinberg operator with minimal $L$ violation, Majorana neutrino masses   \\
$\dfrac{QQQL}{\Lambda^2}$,   $\dfrac{u^cu^cd^ce^c}{\Lambda^2}$ & 6 & 1 & 1 &
\begin{tabular}[c]{@{}l@{}}
$p \to e^+\pi^0$ \\
$p \to \bar{\nu}K^+$
\end{tabular} &
Characteristic $\Delta B=\Delta L=1$ operators appearing in GUTs, typically $\Delta(B-L)=0$  \\

$\dfrac{(u^cd^cd^c)(u^cd^cd^c)}{\Lambda^5}$ & 9 & 2 & 0 &
\begin{tabular}[c]{@{}c@{}}
$n -\bar n$ \\ 
dinucleon decay 
\end{tabular}
&
Minimal $\Delta B=2$ quark   operators, $\Delta(B-L)=2$  \\

$\dfrac{(\bar{L}_i\sigma^{\mu\nu}e_j)H F_{\mu\nu}}{\Lambda^2}$ & 6 & 0 & 0 &
\begin{tabular}[c]{@{}l@{}}
$\mu\to e\gamma$\\
$\tau\to\mu\gamma$  \end{tabular}
&
CLFV dipole, probes lepton flavor structure while conserving total $L$ \\

$\dfrac{u^cd^cd^c \chi}{\Lambda^2}$ & 6 & 1 & 0 &
\begin{tabular}[c]{@{}c@{}}
$n\to \chi+\gamma$\\ 
$p\to \chi+ {\rm mesons}$  
\end{tabular} &
Interaction with a neutral fermion $\chi$ potentially carrying $B$, missing visible energy  \\

$\dfrac{(QQQL) \phi}{\Lambda^3}$ & 7 & 1 & 1 &
$p \to e^+\pi^0\phi$ &
Proton decay with emission of a light boson $\phi$ (e.g.~Majoron-like, other scalar), missing visible energy  \\

$\dfrac{(L \tilde{H})(L \tilde{H}) \phi}{\Lambda^2}$ & 6 & 0 & 2 &
$0\nu\beta\beta\phi$ &
Majoron-like $0\nu\beta\beta$, connects to $L$-breaking, missing visible energy \\
\Xhline{1.2pt}
\end{tabular}
\caption{Representative effective operators relevant for probing $B$ and $L$ violation. 
Shown are schematic SMEFT operators and illustrative extensions involving additional light fields.
Their lowest operator dimensions and representative processes characterized by
$(\Delta B,\Delta L)$ selection rules are shown. Gauge, Lorentz  and flavor indices are suppressed for simplicity.
The unique $d=5$ Weinberg operator with $\Delta L=2$ generating Majorana neutrino masses,
characteristic $d=6$ operators with $\Delta B=\Delta L=1$ mediating nucleon decay,
the minimal $d=9$ six-quark operators with $\Delta B=2$ relevant for $n - \bar n$ oscillations,
as well as selected examples involving CLFV
or light new states are included.  Here $F_{\mu\nu}$ is shown schematically and denotes the physical electromagnetic field strength
appearing after electroweak symmetry breaking. Indexes $i,j$ denote lepton flavor  generation.
Operators with additional fermion $\chi$ and scalar $\phi$ showcase EFT extensions beyond minimal SMEFT.}
\label{tab:operator_ladder}
\end{table}

At dimension $d = 6$, operators violating $B$ and $L$  by one unit each, with $\Delta B=1$, $\Delta L=1$ and $|\Delta(B-L)|=0$, first arise through four-fermion interactions~\cite{Wilczek:1979hc}   
  \begin{align}
  \mathcal{O}_6 \sim \dfrac{1}{\Lambda^2} (QQQL),\qquad \mathcal{O}_6 \sim \dfrac{1}{\Lambda^2} (u^c u^c d^c e^c)~.
  \label{eq:dim6BNV_revised}
  \end{align}
 These operators underlie representative nucleon decay channels such as $p \to e^+ \pi^0$ and $p \to \bar{\nu} K^+$, where a proton $(uud)$ decays into a positron or antineutrino plus mesons. 
Such operators can arise from integrating out heavy states that mediate quark-lepton transitions, including heavy gauge bosons or   leptoquarks in UV completions. They are a generic feature of many GUTs, where quarks and leptons reside in common multiplets and naturally lead to possible $\Delta B = \Delta L = 1$ interactions at low energies.
A schematic illustration of such $\Delta B=\Delta L=1$ transitions mediated by heavy bosons in scenarios of unification is shown in Fig.~\ref{fig:processes} for the representative channel $p\to e^+\pi^0$.

At dimension $d = 9$, the lowest dimension operators violating $B$ by two units, with
$\Delta B=2$, $\Delta L=0$ with $|\Delta(B-L)|=2$, are six-quark operators~\cite{Mohapatra:1980qe,Rao:1982gt} of schematic form
  \begin{align}
  \mathcal{O}_9 \sim \dfrac{1}{\Lambda^5} (u^c d^c d^c)(u^c d^c d^c).
  \label{eq:d9nnbar}
  \end{align}
These induce processes such as $n - \bar{n}$  oscillations where a neutron $(udd)$ transforms into an antineutron ($\bar{u} \bar{d} \bar{d}$), as well as dinucleon decays. At the nucleon level  $n - \bar{n}$ oscillations are commonly parameterized by a mass mixing term $\delta m_{n\bar n}$ obtained after matching the quark level operators onto hadronic degrees of freedom.
In UV completions, such operators can arise from integrating out heavy states that generate $\Delta B = 2$ interactions at the quark level. For example, such processes can be mediated by multi-scalar exchange, heavy diquark type contributions  or arising from mechanisms associated with $B - L $   breaking in left-right symmetric scenarios.  
A schematic illustration of such $\Delta B=2$ transitions exemplified by $n - \bar{n}$ oscillations is shown in Fig.~\ref{fig:processes}.

These examples demonstrate that distinct experimental probes are sensitive to operators of different mass dimension and symmetry structure. Hence, they can access distinct ranges of effective energy scales of new physics depending on the underlying Wilson coefficients as well as hadronic and nuclear contributions.
Tab.~\ref{tab:operator_ladder} summarizes representative SM gauge invariant effective operators that violate $B$ and $L$, organized by operator dimension and selection rules $(\Delta B, \Delta L)$ and illustrates how qualitatively distinct experimental searches probe different symmetry structures and energy scales.
Also shown are examples of operators involving additional light new fields, such as a scalar $\phi$ and a fermion $\chi$ that are assumed to be singlets under the SM gauge group, illustrating EFT extensions beyond the SM~\cite{Fridell:2023tpb}.

The Wilson coefficients in Eq.~\eqref{eq:SMEFTexp} are defined by matching to the UV theory at a scale typically of order $\mu\sim\Lambda$ and must
be evolved to lower energy scales using renormalization group equations. Multiple heavy mass thresholds may be present  depending on the underlying UV completion, with different degrees of freedom becoming dynamical or decoupling at different energy scales. In particular, renormalization group evolution, often dominated by QCD strong interactions below the electroweak scale, can significantly modify
the relative strengths of operator structures and thus must be included when translating experimental
limits into constraints on $\Lambda$.
At scale $\mu\sim\mathcal{O}({\rm GeV})$, quark level operators need to be matched onto hadronic degrees of freedom.
For proton decay  the relevant hadronic transition amplitudes are matrix elements of $B$-violating quark operators between an initial proton state $|p\rangle$ and a final meson and lepton state $|M, l \rangle$, schematically given by
\begin{align}
\langle M,\ell|\mathcal{O}_{\Delta B}|p\rangle \sim \sum_k \alpha_k^{(M)}\bar{\ell}\Gamma_k p 
\label{eq:hadronicME}
\end{align} 
where $\mathcal{O}_{\Delta B}$ denotes an effective operator that violates $B$, $M$ is a pseudoscalar meson, $\ell$ a charged lepton or antineutrino,   $\Gamma_k$ denote Dirac structure factors and hadronic coefficients $\alpha_k^{(M)}$ encode the non-perturbative QCD dynamics and depend on the operator basis. 
These hadronic matrix elements could be computed from first principles, such as using numerical simulations of QCD on a discretized space-time lattice (i.e. lattice QCD)~\cite{Aoki:2008ku,Yoo:2021gql}, leading to a significant reduction of theoretical uncertainties in nucleon decay rate predictions.

For physical processes occurring within nuclei, including $0\nu\beta\beta$, $n-\bar{n}$ and
multi-nucleon decays, hadronic operators must be embedded into nuclear many-body systems. This introduces nuclear matrix
elements that encode nuclear structure, correlations  and multi-nucleon dynamics~\cite{Engel:2016xgb}.
For $0\nu\beta\beta$, the decay rate is often expressed, considering a single dominant contributing mechanism, as
\begin{align}
\left(T^{0\nu}_{1/2}\right)^{-1}
 = 
G^{0\nu} |M^{0\nu}|^2 |\eta|^2 ,
\label{eq:0nubb_rate}
\end{align}
where $G^{0\nu}$ is a phase-space factor, $M^{0\nu}$ the nuclear matrix element and $\eta$ a dimensionless parameter containing the
underlying $L$-violating physics.
Short range contributions to $0\nu\beta\beta$ can be systematically incorporated through effective nuclear operators derived within chiral EFT, which provides a systematic framework for separating short distance physics from nuclear dynamics~\cite{Cirigliano:2022oqy}. Systematic classifications of higher dimensional $\Delta L=2$ effective operators beyond the Weinberg operator
have been developed within EFT frameworks~\cite{Babu:2001ex}.
An analogous separation of short distance $B$-violating physics and nuclear responses applies to $\Delta B=2$ processes in nuclei, although the corresponding nuclear operators and response functions differ.

A prominent insight from EFTs is the parametric scaling of decay rates with operator dimension 
\begin{align}
\Gamma  \sim \frac{|C^{(d)}|^2}{\Lambda^{2(d-4)}} |{\cal M}|^2 \times (\text{phase space \& kinematic factors}) ,
\label{eq:ratescale}
\end{align}
where ${\cal M}$ denotes the relevant hadronic or nuclear matrix element.
The experimentally accessible observable is often the lifetime  or half-life of a physical process, which is inversely proportional to the decay rate as $\tau = 1/\Gamma$ and $T_{1/2} = \ln 2/ \Gamma$, respectively. As a result, this highlights that even modest improvements in lifetime limits for a particular physical process can correspond to substantial increases in the effective scale
$\Lambda$ tested by rare event searches, especially for lower dimensional operators.

Although violations of individual lepton flavor number $\Delta L_e$, $\Delta L_{\mu}$ and $\Delta L_{\tau}$ may conserve total $L =  L_e + L_\mu + L_\tau$, they are naturally described within
the same EFT framework when physics beyond SM is considered~\cite{deGouvea:2013zba}.
As an illustrative example, CLFV can be generated in the SMEFT by higher dimensional, gauge invariant operators. After electroweak symmetry breaking and matching onto electromagnetic dipole operators they can induce processes such as $\mu\to e\gamma$. Hence, lepton flavor violation processes provide
complementary information to $B$ and $L$ on the underlying structure and characteristic scales of physics beyond SM.

\subsection{Fundamental theory and unification}

The pursuit of a unified description of fundamental interactions has been a longstanding motivation for theories beyond SM, in which $B$ and $L$ conservation emerge as accidental symmetries at low energies.
The $B$ and $L$ global symmetries are not imposed by gauge invariance in SM and can be expected to be violated in a more fundamental framework. Although such theories typically operate at energy scales far beyond the reach of direct probes, such as with particle accelerators, they often predict rare processes that can be tested experimentally.

GUTs provide a prominent realization of this concept by embedding SM gauge
group $SU(3)_C \times SU(2)_L \times U(1)_Y$ into a single group $G_{\rm GUT}$ at a high
unification scale $M_{\rm GUT}$ often $\sim 10^{15}-10^{16}$~GeV.
These theories can motivate charge quantization and could lead to approximate gauge coupling unification. Here, SM quarks and leptons are assigned to common gauge group multiplets. This enables interactions and quark-lepton transitions that violate $B$ and $L$ through the exchange of new heavy gauge or scalar fields. 
A canonical illustration is the Georgi-Glashow $SU(5)$ model~\cite{Georgi:1974sy}.
In the minimal $SU(5)$ the SM fermions of each generation are unified into   irreducible gauge group 
representations $\overline{\mathbf{5}} = (d^c,  L)$, $\mathbf{10} = (Q,  u^c,  e^c)$. This unification of quarks and leptons within joint multiplets and the corresponding $SU(5)$ generators imply the existence of
heavy gauge bosons $X$ and $Y$, transforming as $(\mathbf{3},\mathbf{2})_{-5/6}$ and its conjugate under  
$SU(3)_C \times SU(2)_L \times U(1)_Y$.
These gauge bosons can mediate quark-lepton interactions, with $ 
u + u \to  X \to e^+ + \bar{d}$ as schematic illustration. At energies well below $M_{\rm GUT}$, with heavy mediators integrated out, these effects are described by effective $d=6$ operators~\cite{Wilczek:1979hc}
\begin{equation}
\mathcal{L}_{\Delta B, \Delta L}
 \supset 
\frac{1}{M_{\rm GUT}^2}
\left( \bar{q} \gamma^\mu q \right)
\left( \bar{q} \gamma_\mu \ell \right)
+ \text{h.c.},
\label{eq:gutd6}
\end{equation}
where for simplicity the heavy gauge boson masses have been identified with the unification scale 
$M_X \sim M_{\rm GUT}$  and the interaction is stated in four component Dirac notation. This leads to nucleon decay such as $p \to e^+ \pi^0$, as depicted in Fig.~\ref{fig:processes},
with $\Delta B = \Delta L = 1$ and 
conserving $B-L$  for this operator class.

The general EFT scaling of decay rates with operator dimension of
Eq.~\eqref{eq:ratescale} well illustrates the concepts in the case of $B$-violating $d = 6$ operators inducing nucleon decay such as Eq.~\eqref{eq:gutd6}.
Neglecting prefactor coefficients and hadronic uncertainties, the proton
lifetime scales parametrically as
\begin{equation}
\tau_p \sim \frac{\Lambda^4}{m_p^5},
\label{eq:tauScaling}
\end{equation}
where $\Lambda$ denotes the characteristic scale suppressing the effective
operator and $m_p = 0.938$~GeV is the proton mass.
For a unification scale value of
$\Lambda \sim M_{\rm GUT} \sim 10^{16}~\mathrm{GeV}$ yields proton lifetimes $\tau_p \gg 10^{30}$~years that is
many orders of magnitude beyond typical laboratory timescales. This showcases why testing such
processes require extremely large and long running experiments.  
The estimates here are intended only as parametric illustrations. Detailed lifetime predictions
depend sensitively on broad range of considerations such as hadronic matrix elements, flavor structure, details of microscopic theory at high energies and can vary by multiple
orders of magnitude across theoretical models~\cite{Langacker:1980js,Nath:2006ut}.

Partial unification schemes such as the Pati-Salam model~\cite{Pati:1973uk}, based on
$SU(4)_C \times SU(2)_L \times SU(2)_R$,
unify quarks and leptons by treating leptons as a fourth color. While devoid of some features of minimal GUTs, these models also predict $B$-violating and $L$-violating interactions. They are typically mediated by new leptoquark states that enable lepton-quark transformations, often with distinctive selection rules and resulting nucleon decay signatures.

Left-right symmetric theories extend the SM gauge group to
$SU(3)_C \times SU(2)_L \times SU(2)_R \times U(1)_{B-L}$,
restoring parity symmetry at high energies and promoting $B-L$ to a local gauge symmetry~\cite{Mohapatra:1974gc,Senjanovic:1975rk}. The spontaneous breaking of this symmetry requires the presence of right handed neutrino fields $N_R$, which are singlets under the SM gauge
group.
A Majorana mass term can then be generated 
\begin{equation}
\mathcal{L}_M
=
-\frac{1}{2} M_N   \overline{N_R^c} N_R + \text{h.c.},
\end{equation}
where $M_N$ denotes the Majorana mass scale associated with the breaking of
$B-L$.
This term induces $\Delta L = 2$, and together with electroweak scale
Dirac neutrino masses leads to neutrino masses that are naturally light through the seesaw mechanism~\cite{Minkowski:1977sc,Gell-Mann:1979vob,Yanagida:1979as}.  From the low energy perspective, such constructions provide a concrete UV completion of the unique dimension five $\Delta L = 2$ Weinberg operator of Eq.~\eqref{eq:weinbergop}. This illustrates how $L$ violation encoded in effective operators can emerge from extended underlying structure.

A more comprehensive framework is provided by $SO(10)$ GUT~\cite{Fritzsch:1974nn},
although alternative unified structures and embeddings are also possible
\cite{Langacker:1980js}. The $SO(10)$ contains the
$SU(5)$ and Pati-Salam gauge structures as subgroups, enabling to establish a common
group-theoretic origin of their characteristic features.
At the level of fermion representations a single $SO(10)$ irreducible representation multiplet $\mathbf{16}$ decomposes under $SU(5)$ as
$\mathbf{16} = \mathbf{10} \oplus \overline{\mathbf{5}} \oplus \mathbf{1}$,
where the singlet $\mathbf{1}$ corresponds to a right handed neutrino.
Hence, $SO(10)$ GUT unifies SM fermion generation and a right handed neutrino and its embedding automatically extends minimal $SU(5)$.
Under the Pati-Salam subgroup
$SU(4)_C \times SU(2)_L \times SU(2)_R$ the same $\mathbf{16}$ decomposes as
$(\mathbf{4},\mathbf{2},\mathbf{1}) \oplus (\overline{\mathbf{4}},\mathbf{1},\mathbf{2})$,
where leptons appear as a fourth color and right-handed fermions transform as
$SU(2)_R$ doublets.
Hence, $B-L$ arises as a generator of $SU(4)_C$, making it manifest
that $B$ and $L$ are unified quantum numbers whose violation is
linked to symmetry breaking.

From an experimental perspective, unified frameworks such as $SO(10)$ highlight that distinct
low energy signatures, such as  nucleon decay with $\Delta B = 1$,  $n - \bar{n}$ oscillations with $\Delta B
= 2$ and $\Delta L = 2$ processes associated with
Majorana neutrino masses, can originate from a single underlying
structure. 
Such theories can naturally accommodate neutrino masses and permit both $\Delta(B-L)=0$ and
$\Delta(B-L)=2$ processes, establishing connections between $B$ violation, $L$ violation and potentially the origin of the cosmic matter-antimatter asymmetry.
Importantly, the relative pattern of observable processes and their rates is governed by
selection rules arising from symmetry-breaking structures and the mass spectrum of heavy
mediator fields.
This highlights that the resulting experimental searches that probe complementary $(\Delta B,\Delta L)$ channels
provide a powerful means of discriminating between different classes of underlying theories,
even in the absence of direct access to the relevant high energy scales.

SUSY frameworks postulate a symmetry between integer spin and half-integer spin
quantum fields, extending the SM such that each particle is accompanied by a superpartner
with spin differing by one half.
Beyond addressing the gauge hierarchy problem, SUSY improves precision of gauge coupling unification at high scales,
particularly within context of SUSY GUTs~\cite{Dimopoulos:1981yj}.
When SUSY is promoted to a local symmetry gravity is naturally incorporated through
supergravity~\cite{Wess:1992cp}, which is of interest for theories of quantum gravity and string theory.
In generic SUSY SM extensions $B$-violating and $L$-violating interactions
are allowed by symmetry structure unless additional restrictions are imposed.
At the renormalizable level, $d = 4$ operators can induce rapid nucleon decay.
Discrete symmetries such as $R$-parity~\cite{Farrar:1978xj} or matter parity~\cite{Dimopoulos:1981zb}
are commonly invoked to forbid these operators.

Even when  such renormalizable $B$-violating and $L$-violating interactions are eliminated,
SUSY GUTs often predict $B$ violation and $L$ violation through higher dimensional operators.
Such violations can arise at dimension $d = 5$ in the superpotential~\cite{Sakai:1981pk,Weinberg:1981wj,Hisano:1992jj} 
\begin{equation}
W_{\Delta B,\Delta L}\supset
\frac{1}{M_{\rm GUT}}
\left( Q Q Q L + U^c U^c D^c E^c \right),
\label{eq:SUSYdim5}
\end{equation}
where $Q$, $L$, $U^c$, $D^c$, and $E^c$ here denote chiral superfield multiplets and contain both fermionic
fields and their scalar superpartner components.
In component form  these dimension $d = 5$ superpotential operators generate interactions involving two fermions and two scalar superpartners, rather than four-fermion contact terms of dimension $d = 6$, which distinguishes them from the non-SUSY scenario. The four-fermion operators relevant for nucleon decay into SM particles arise only after additional dressing by SUSY superpartner gaugino or higgsino exchange, introducing a dependence on the superpartner mass spectrum associated with SUSY breaking.
Notably, such operators can dominate nucleon decay amplitudes and lead to decay modes distinct from those
in non-SUSY theories, such as $p \rightarrow \bar{\nu} K^+$.
Their effective suppression depends sensitively on the UV completion and on the symmetries
governing $B$ and $L$ at lower energies.

This illustrates a general principle that null results in nucleon decay and other rare process experimental searches do not necessarily imply that such processes are suppressed by extremely high mass scales of a more fundamental theory. Instead, they could reflect symmetry-based selection rules that forbid the lowest dimensional operators and related physical processes altogether.
Further examples include discrete $\mathbb{Z}_N$ symmetries~\cite{Ibanez:1991pr,Dreiner:2005rd,Chen:2014gua} that can eliminate lower dimensional operators while permitting higher dimensional ones or alternative $B$-violating and $L$-violating processes. In such scenarios, the dominant experimental signatures could shift toward typically suppressed multi-body nucleon decays or channels involving heavier mesons.

In theories with extra spatial dimensions, $B$ violation and $L$ violation can be strongly
influenced by the geometric localization of fermion fields, providing an additional organizing
principle that does not rely on imposing new symmetries explicitly.
If quarks and leptons are localized at different positions in the extra dimensions, effective
four dimensional $B$-violating and $L$-violating operators can be exponentially suppressed
by small wavefunction overlaps, even when such operators are allowed by the symmetries of the
underlying higher dimensional theory~\cite{Arkani-Hamed:1999ylh}. At the level of the resulting four dimensional EFT, a generic operator inducing
$\Delta B = \Delta L = 1$ transitions may appear schematically as
\begin{equation}
\mathcal{L}_{\Delta B, \Delta L}
\supset
\frac{c}{M_*^{2}}  
(qqq\ell)  +  \text{h.c.},
\end{equation}
where $q$ and $l$ denote generic quark and lepton fields, with gauge, Lorentz  and flavor contractions being implicit, $M_*$ denotes the characteristic fundamental scale of the higher dimensional theory and the coefficient $c$ encodes geometric suppression due to fermion localization in the extra dimensions.
Analogous considerations apply to higher dimensional operators such as $\Delta B = 2$ interactions inducing $n - \overline{n}$, which can remain experimentally accessible even when proton
decay is significantly suppressed~\cite{Nussinov:2001rb}.
More generally, extra-dimensional constructions can generate correlated or hierarchical patterns
of $B$ violation and $L$ violation considering spacetime structure itself.

Finally, general considerations from quantum gravity suggest that exact global symmetries do not exist in nature~\cite{Banks:2010zn,Harlow:2018tng}. In the presence of black holes or Planck scale dynamics global charges such as  $B$
and  $L$ are thus expected to be violated, giving rise to higher dimensional effective
operators suppressed by powers of the reduced Planck mass
$M_{\rm Pl}  \simeq 2.4\times 10^{18}$~GeV, or by related UV scales characterizing quantum-gravitational effects. From an EFT perspective, such effects can be parameterized schematically as
\begin{equation}
\mathcal{L}_{\rm eff}
 \supset 
\sum_{d>4} \frac{c_d}{M_{\rm Pl}^{ d-4}} \mathcal{O}_d ,
\end{equation}
where $\mathcal{O}_d$ denotes operators that violate $B$ or $L$ and the coefficients $c_d$ encode unknown microscopic UV physics.
Notably, nucleon stability is not automatically guaranteed even in frameworks regarded as UV complete, such as string theory or brane world constructions. This often requires additional structure, such as gauged $B - L$, discrete gauge symmetries or geometric suppression mechanisms, to sufficiently suppress $B$-violating operators~\cite{Nath:2006ut}. Thus, searches for nucleon decays can also be viewed as probes of physics at extremely
high mass scales and potentially even approaching the Planck scale, as emphasized in
Ref.~\cite{Harnik:2004yp}.
While exact global symmetries are expected to be violated by quantum gravity effects at some level, certain discrete symmetries may remain exact, allowing $B$ violation and $L$ violation to be highly suppressed without being strictly forbidden. These considerations provide a powerful, largely model-independent motivation for broad experimental searches of $B$ violation and $L$ violation.

This highlights that $B$ violation and $L$ violation emerges not as unusual anomalies, but as a generic consequence of deeper principles governing fundamental interactions. Experimental searches for such processes can therefore probe physics at energy scales far beyond
the direct reach of techniques such as particle accelerators, providing a unique window into fundamental
theory.

\subsection{Cosmological significance of baryon and lepton number violation}
\label{subsec:cosmology}

One of the strongest motivations for $B$ violation and $L$ violation arises from
cosmology.
Observations of the cosmic microwave background radiation and primordial light element abundances establish
a small but non-zero matter-antimatter asymmetry in the Universe. This is conventionally expressed as the
baryon-to-entropy ratio~\cite{ParticleDataGroup:2024cfk}
\begin{equation}
\eta_B  =  \frac{n_B-n_{\bar B}}{s}  \simeq  10^{-10},
\end{equation}
where $n_B$ ($n_{\bar B}$) is the baryon (antibaryon) number density  and $s$ is the entropy density, which is proportional to the photon number density after considering entropy conservation.
This asymmetry is challenging to explain within the minimal SM under standard cosmological assumptions.
Any successful dynamical explanation of the baryon asymmetry, referred to as baryogenesis,
must satisfy the three Sakharov conditions~\cite{Sakharov:1967dj}:
(i) violation of $B$,
(ii) violation of $C$ and $CP$ symmetries,
and (iii) a departure from thermal equilibrium.
These conditions are general and largely model-independent. As a consequence, $B$ violation plays a central role in known dynamical explanations
of the observed dominance of matter over antimatter.

In the SM, non-perturbative electroweak effects violate $B+L$ through sphaleron transitions while
preserving $B-L$.
At the effective operator level, these effects are encoded in the non-perturbative instanton interaction~\cite{tHooft:1976up}, schematically
\begin{equation}
\mathcal{O}_{B+L}
 \sim 
\prod_{i=1}^{3}
\left(
q_{Li}  q_{Li}  q_{Li}  \ell_{Li}
\right),
\end{equation}
which involves left handed fermion doublets of the three SM generations  and carries $\Delta B = \Delta L = 3$.
Although exponentially suppressed at zero temperature, such processes are active in the early
Universe~\cite{Kuzmin:1985mm} and play a central role in baryogenesis scenarios. Since these processes conserve $B-L$  they cannot generate a net $B-L$ asymmetry from symmetric initial conditions, instead
they can convert pre-existing $B-L$ asymmetry into correlated $B$ and $L$ asymmetries.
This highlights the significance of either explicit $B$ violation beyond the
SM or $L$ violation whose asymmetry can be partially converted into $B$ by
sphaleron processes. From a cosmological perspective, the observed matter-antimatter asymmetry elevates $B$ violation and $L$ violation to central features of fundamental
physics.
Experimental probes of $\Delta B\neq0$ and $\Delta L\neq0$ processes therefore test not only
specific SM extensions, but also aspects of the frameworks
invoked to explain the origin of the cosmic matter-antimatter asymmetry.

A prominent class of scenarios exploiting $L$ violation to generate the
cosmic baryon asymmetry is leptogenesis~\cite{Fukugita:1986hr}.
Here, an initial lepton asymmetry produced at high temperatures is reprocessed into a
baryon asymmetry by sphaleron interactions.
One representative realization arises in seesaw frameworks with heavy Majorana fermions~\cite{Minkowski:1977sc,Gell-Mann:1979vob,Yanagida:1979as}, for which the
light neutrino masses are effectively generated via
\begin{equation}
m_\nu  \simeq  -   m_D   M^{-1}   m_D^T ,
\end{equation}
where $m_D$ denotes the Dirac neutrino mass matrix and $M$ the Majorana mass matrix of heavy right handed (SM-singlet, sterile) neutrino states.
Out-of-equilibrium, $CP$-violating decays of these heavy right handed neutrino particles generate a lepton asymmetry,
which is subsequently converted into baryon number according to $B  =  c_{\rm sph}  (B - L )$ in thermal equilibrium, with $c_{\rm sph}  =  28/79$ 
for the SM particle content~\cite{Harvey:1990qw,Buchmuller:2005eh}.
The conversion factor is fixed by equilibrium conditions and the anomaly structure, illustrating
the connection between $B$ violation and $L$ violation in the early Universe.

$L$-violating searches are complementary to other approaches that probe the absolute scale of neutrino masses. In addition to kinematic measurements, such as in beta decay, and cosmological constraints inferred from the large scale evolution of the Universe, proposed experiments aiming at the direct detection of the cosmic neutrino background, such as through neutrino capture on beta unstable targets as in PTOLEMY~\cite{PTOLEMY:2018jst}, would in principle provide access to the neutrino mass scale through a qualitatively distinct observable. These considerations highlight the broad interplay between laboratory, astrophysical  and cosmological probes in constraining neutrino properties and the origin of $L$ violation, independent of the specific UV realization.

Nontrivial but indirect connections exist  between cosmology and low energy experimental searches
for $B$ violation and $L$ violation.
$L$-violating interactions that are active in thermal equilibrium at sufficiently high
temperatures can, in some scenarios, erase a pre-existing asymmetry through washout effects. Thus, laboratory searches for $L$ violation can constrain certain classes of high scale
baryogenesis setups.
In GUT frameworks, $B$ violation and $L$ violation are often internal features
rather than additional constituents. Thus, interactions in such theories often generate both $\Delta B\neq0$
and $\Delta L\neq0$ processes, and baryogenesis might proceed through the out-of-equilibrium decays
of heavy GUT states. It is then feasible that
baryogenesis, neutrino mass generation and nucleon decays can arise from common underlying framework and thus cosmological aspects can inform laboratory searches and vice versa. If $B$ or $L$ violation is associated with symmetry breaking phase transitions in the early Universe, observable phenomena such as stochastic gravitational wave backgrounds may appear~\cite{Caprini:2018mtu}, as illustrated in Ref.~\cite{King:2020hyd}. 
While highly model-dependent, these considerations illustrate how cosmological observations can
complement laboratory searches for $B$ and $L$  violation.

On much longer timescales than the current age of the Universe, $B$ violation could also influence the ultimate fate of
astrophysical structures and the Universe itself.
After conventional stellar evolution ceases and nuclear fusion is no longer viable as an
energy source, the stability of ordinary (baryonic) matter becomes a limiting factor for the persistence of
stars and compact object remnants such as neutron stars and white dwarfs.
As analyzed in studies of the futuristic evolution of the Universe~\cite{Adams:1996xe}  even considering for illustration
extremely slow nucleon decay with lifetimes of $\tau_p\sim10^{34}-10^{36}$~years, as predicted by some theories, would
eventually dominate energy release from baryons.
For a solar mass $M_{\odot}$ object  the corresponding power from proton decay is of order
$L\sim M_{\odot}/\tau_p \sim 10^{11}-10^{13}$~erg~s$^{-1}$ that is many orders of magnitude below the present
solar luminosity, but could sustain very low luminosity emission over timescales significantly
exceeding those of stellar fusion.
Although such epochs lie far beyond direct observational reach, this illustrates that $B$ violation is not only relevant to unification and the origin of matter, but may also
govern the ultimate persistence of baryons and the long-term fate of the visible Universe.

\section{Experimental Tests of Baryon Number Conservation}

\subsection{Nucleon decays ($\Delta B=1$)}

Direct searches for nucleon decays provide the most straightforward experimental probes of $B$ violation. In a broad range of motivated SM extensions, such as GUTs, $\Delta B = 1$ nucleon decay processes can arise from EFT operators suppressed by a high mass scale that is typically associated with unification. An observation of nucleon decay would therefore constitute direct evidence for new physics and  offer a unique window into the structure of more fundamental interactions at energy scales challenging to probe with other techniques.

Already prior to dedicated experiments,  Goldhaber noted that if protons decayed with lifetimes many orders of magnitude shorter than those known, the associated energy deposition from decay products in ordinary matter could lead to biologically observable radiation effects~\cite{Goldhaber:1980dn}. This provided early qualitative arguments for the observed stability of matter. The earliest direct experimental searches for nucleon decays were carried out by Reines, Cowan, Goldhaber in 1954, which looked for evidence of proton instability
through the detection of anomalous energy deposition in large scintillators and set limits on proton lifetime of around $\sim 10^{22}$~years~\cite{Reines:1954pg}.

\begin{figure}[t]
\centering
\includegraphics[width=0.65\linewidth]{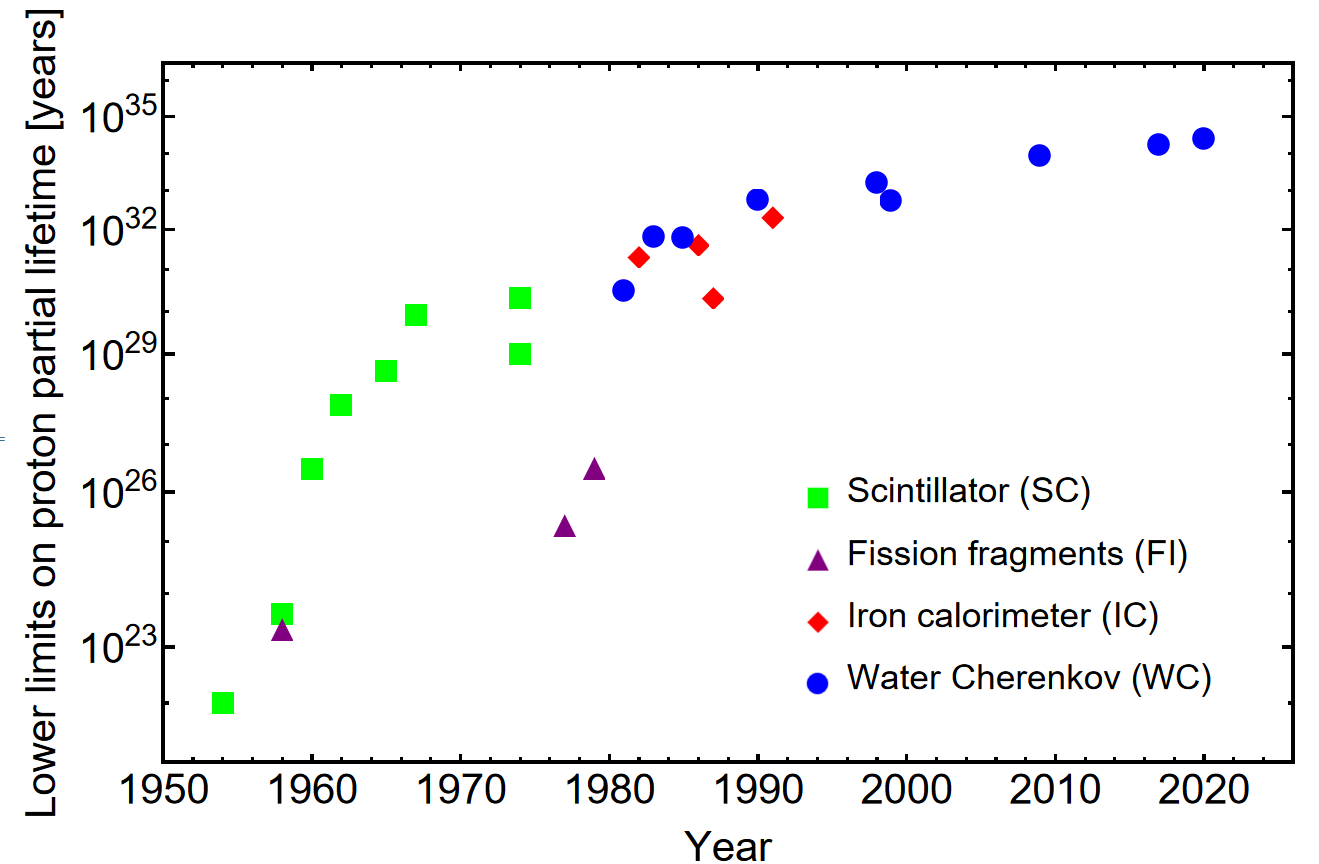} 
\caption{Historical evolution of experimental lower limits on the proton partial lifetime, primarily corresponding to the benchmark decay channel $p \rightarrow e^+ \pi^0$. Earlier results based on more general or distinct searches are included to illustrate the development of techniques. 
Detection principles are denoted as follows: liquid scintillator (SC, Ref.~\cite{Reines:1954pg,Reines:1958pf, Backenstoss1960,Giamati:1962pe,Kropp:1965pd, Gurr:1967pc, Bergamasco1974, Reines:1974pb}),  searches for fission fragments from radioactive ore (FI, using $^{232}$Th~\cite{Flerov:1958zz}, $^{130}$Te~\cite{EvansJr:1977zuj}, $^{39}$K~\cite{Fireman:1979xr}),
and fine-grained calorimeters primarily employing iron $^{26}$Fe (IC, from NUSEX~\cite{Battistoni:1982vv}, KGF~\cite{Krishnaswamy:1985tho}, Soudan~\cite{Soudan-1:1986zji}, Frejus~\cite{Frejus:1990myz} experiments), water Cherenkov (WC, from Homestake~\cite{Cherry:1981uq}, IMB~\cite{Irvine-Michigan-Brookhaven:1983iap,Gajewski:1989gh,McGrew:1999nd}, Kamiokande~\cite{Arisaka:1985lki}, Super-Kamiokande~\cite{Super-Kamiokande:1998mae,Super-Kamiokande:2009yit,Super-Kamiokande:2016exg,Super-Kamiokande:2020wjk} experiments). A confirmed detection signal would manifest as a finite measured lifetime, and the historical record contextualizes sensitivity evolution up to discovery.
}
\label{fig:pdkhist}
\end{figure}

Subsequent experimental searches progressively improved
sensitivity over the following decades and with a range of complementary techniques.
Fig.~\ref{fig:pdkhist} summarizes the historical evolution of experimental lower limits on the proton partial lifetime, primarily
corresponding to the benchmark decay channel $p \rightarrow e^{+}\pi^{0}$. This illustrates the steady improvements achieved through increased detector observational exposure, improved background rejection and advances in detector technology.
Early technologies
included scintillators~\cite{Reines:1954pg,Reines:1958pf, Backenstoss1960,Giamati:1962pe,Kropp:1965pd, Gurr:1967pc, Bergamasco1974, Reines:1974pb}, radiochemical techniques searching for fission fragments from radioactive ore~\cite{Flerov:1958zz,EvansJr:1977zuj,Fireman:1979xr} and fine-grained calorimeters primarily made of iron including NUSEX~\cite{Battistoni:1982vv}, KGF~\cite{Krishnaswamy:1985tho}, Soudan~\cite{Soudan-1:1986zji} and Frejus~\cite{Frejus:1990myz} experiments,
with sensitivities limited by relatively modest target masses and higher backgrounds.   Over subsequent decades, progressively larger underground detectors improved lifetime limits by orders of magnitude. Modern efforts have been spearheaded by water Cherenkov experiments, including 
Homestake~\cite{Cherry:1981uq}, Irvine-Michigan-Brookhaven (IMB)~\cite{Irvine-Michigan-Brookhaven:1983iap,Gajewski:1989gh,McGrew:1999nd}, Kamiokande~\cite{Arisaka:1985lki}, Super-Kamiokande~\cite{Super-Kamiokande:1998mae,Super-Kamiokande:2009yit,Super-Kamiokande:2016exg,Super-Kamiokande:2020wjk}. Super-Kamiokande, building on the successful realization of earlier underground experiments, has played a defining role in establishing benchmark limits on a broad range of  nucleon decay channels. 
Subsequent generations of large water Cherenkov detectors, such as Hyper-Kamiokande~\cite{Hyper-Kamiokande:2018ofw}, can further enhance sensitivity to nucleon decay processes through substantially increased exposure. Such large experiments are multipurpose instruments, capable of exploring a broad range of physics beyond just nucleon decays.

While some early experiments reported candidate nucleon decay events these typically occurred in background-limited searches 
and did not constitute statistically significant excesses over expectation. None were subsequently confirmed by later 
observations with increased experimental exposure and improved background control. Modern experimental searches for nucleon decay are driven by the empirical finding that nucleon lifetimes exceed $\mathcal{O}(10^{30})$~years as well as by theoretical expectations that viable scenarios may reside at still higher scales. This necessitates detectors with large target masses, low backgrounds  and long-term stability.

For such rare event searches requiring high sensitivity, underground detector positioning enables significant background suppression, especially from cosmic ray related processes. 
Across detector technologies  cosmic ray-induced atmospheric neutrino interactions constitute a dominant background class  and in many channels provide an effectively irreducible contribution for nucleon decay searches~\cite{Honda:2015fha}. Efficient background suppression often relies on a combination of factors such as signal event topology, kinematic reconstruction, timing information and coincidence techniques. Systematic uncertainties associated 
with various aspects of the searches such as
nuclear effects, hadronic final-state interactions  and atmospheric neutrino interaction modeling are incorporated into modern analyses. Since nucleon decay searches often   probe bound nucleons, nuclear binding effects, Fermi motion, intranuclear 
re-scatterings  and nuclear de-excitations can modify the observable final state kinematics and thus introduce persistent systematic uncertainties.

 A representative example is $p \rightarrow e^{+}\pi^{0}$, predicted to be dominant in many GUTs~\cite{Langacker:1980js}. With $\pi^0\to 2\gamma$ decays this mode produces characteristic fully visible final state allowing nearly complete kinematic reconstruction of the parent proton's invariant mass and momentum. This mode has long served among primary experimental $B$ violation search benchmarks. With searches constraining the partial lifetime in this channel to the level of $\mathcal{O}(10^{34})$~years and beyond~\cite{Super-Kamiokande:2020wjk}, these results place strong pressure on minimal realizations of unification frameworks and exclude some of them. Similar modes such as $p \rightarrow \mu^+\pi^0$, which can be significant in some models like flipped $SU(5)$~\cite{Ellis:2002vk}, have also been extensively analyzed~\cite{Super-Kamiokande:2020wjk}. Overview of early Super-Kamiokande nucleon decay analyses and methodologies can be found in Ref.~\cite{Takhistov:2016eqm}.

A large class of motivated scenarios, especially based on SUSY GUTs, predict enhanced rates for decay channels such as $p \rightarrow \bar{\nu} K^{+}$~\cite{Sakai:1981pk,Weinberg:1981wj}. Unlike nucleon decay channels with fully visible final states occurring inside the detector, 
neutrinos produced in $p \rightarrow \bar{\nu} K^{+}$ decays have negligible interaction rates 
and therefore escape undetected (here, $\nu$ and $\bar{\nu}$ are not being distinguished).
Signal observation thus relies on possible detection of   kaon and its decay products, such as delayed muons from $K^{+} \rightarrow \mu^{+}\nu$ and charged pions. In nucleon decay searches without fully visible final states prominent role can also be played by associated nuclear de-excitation signatures. Depending on the decay channel and parent nucleus, the emission of low energy photons, neutrons or $\alpha$ particles from the excited residual nucleus is possible~\cite{Ejiri:1993rh,Kamyshkov:2002wp}. Current searches constrain the partial lifetime for $p \rightarrow \bar{\nu} K^{+}$ to exceed $\mathcal{O}(10^{33})$~years~\cite{Super-Kamiokande:2014otb}, placing strong pressure on 
minimal realizations of unified theories that predict enhanced strange meson final states. These searches highlight the significance of sensitivity to final state particles with energies at or below the GeV scale, 
characteristic of decaying nucleons at rest.

Beyond two body decay channels, variety of processes with visible final states have also been analyzed in multi-body 
nucleon decay channels. These include fully visible final states such as three charged leptons $l^{\pm} = e^{\pm}, \mu^{\pm}$ in $p \rightarrow l^{+}l^{-}l^{+}$~\cite{Super-Kamiokande:2020tor}, which can be prominent in models such as those involving flavor symmetries~\cite{Hambye:2017qix},
and 
decays involving multiple mesons in the final states~\cite{McGrew:1999nd}. Although such channels are typically phase-space suppressed 
relative to two body modes, their fully reconstructible final states with more complex signal topology permit efficient background rejection in large underground detectors.   

\subsection{Complementarity of detection techniques}

Searches for $B$ violation rely on a diverse array of detection techniques that are advantageous for different classes of final states, background rejection strategies and systematic uncertainties. They include large underground detectors employing distinct target materials and readouts such as water Cherenkov, liquid scintillator  and liquid argon time projection chamber (LArTPC) technologies. The  complementarity of these approaches arises from differing sensitivity to signal event topology and kinematics, energy thresholds, nuclear effects in bound nucleon decays and relevant backgrounds, enabling broad coverage of plausible signatures.

An additional conceptual distinction arises between decays of free protons and  nucleons
bound in nuclei. A notable example is  hydrogen  nuclei in water. Decays of such protons are free from 
nuclear effects resulting in  clean and 
well defined final state kinematics.  Thus, free protons also enable more straightforward interpretations of potential discovery observations.
However, predominantly, nucleon decay searches  probe nucleons bound in nuclei. Nuclear effects such as binding energy and Fermi motion, possible correlations among nucleons including multi-nucleon dynamics~\cite{Yamazaki:1999gz}, and final state interactions of hadrons in the nuclear medium (e.g.  meson re-scattering, absorption) modify observable kinematics and signal efficiencies, constituting prominent irreducible systematic uncertainties. Modern analyses incorporate these effects through dedicated nuclear modeling and detailed Monte Carlo simulations~\cite{Hayato:2002sd}.In simulations of rare nucleon decay searches, final state particle spectra are often generated
using phase-space distributions. While for two-body decays the final state kinematics are uniquely determined, for multi-body channels energy-momentum conservation alone is insufficient to determine spectra and decay dynamics encoded in the matrix element can significantly shape the observable distributions
\cite{Chen:2014ifa}.

Large water Cherenkov detectors, such as IMB~\cite{Irvine-Michigan-Brookhaven:1983iap,Gajewski:1989gh,McGrew:1999nd}, Kamiokande~\cite{Arisaka:1985lki} as well as Super-Kamiokande~\cite{Super-Kamiokande:1998mae,Super-Kamiokande:2009yit,Super-Kamiokande:2016exg,Super-Kamiokande:2020wjk} 
have provided leading sensitivity for many  $\Delta B = 1$ nucleon decay channels due to their scalability to multi-kiloton fiducial detector masses and efficient reconstruction of relativistic charged particles using Cherenkov radiation~\cite{Cherenkov:1934ilx}. Cherenkov emission occurs when a charged particle traverses a medium faster than the speed of light in that medium satisfying $\beta = v/c > 1/n$,  where $v$ is the particle speed, $c$ is the speed of light in vacuum and $n$ is the medium’s index of refraction. The corresponding emission momentum threshold for a particle of mass $m$  
\begin{equation}
p_{\rm thr}=\frac{m}{\sqrt{n^2-1}}~.
\end{equation}
These detectors excel at reconstructing fully visible relativistic final states such as $p\to e^+\pi^0$ through characteristic Cherenkov ring topologies. Particles below Cherenkov threshold  are not observed directly in water and are instead inferred through correlated signatures from decay products, nuclear de-excitation activity  and other secondary observables,
as   illustrated for the case of $K^+$ in $p\to \bar{\nu} K^+$ channel~\cite{Super-Kamiokande:2014otb}. Within the class of Cherenkov detectors, invisible decay searches, in which the primary $B$-violating processes produce 
only weakly interacting or unobservable final states, have also been carried out in deuterium D$_2$O heavy water experiments such as SNO~\cite{SNO:2003lol}. The sensitivity here is provided by nuclear de-excitation and neutron emission in a distinct target nucleus,  rather than from direct reconstruction
of the primary decay products, illustrating complementary nuclear response and systematic coverage relative to large water Cherenkov detectors.

\begin{table}[t]
\centering
\label{tab:BNV_complementarity_matrix}
\begin{tabular}{C{2.5cm}| P{3cm} P{3cm} P{3cm} P{2.5cm}}
\Xhline{1.2pt}
\multicolumn{1}{C{2.5cm}|}{\makecell{\textbf{Detector}\\\textbf{Class}}} &
\multicolumn{1}{>{\centering\arraybackslash}m{3cm}}{\makecell[c]{\textbf{Fully Visible}\\\textbf{Final States}}} &
\multicolumn{1}{>{\centering\arraybackslash}m{3cm}}{\makecell[c]{\textbf{Sub-threshold or}\\\textbf{Delayed Signals}}} &
\multicolumn{1}{>{\centering\arraybackslash}m{3cm}}{\makecell[c]{\textbf{Invisible or Partially}\\\textbf{Invisible Final States}}} &
\multicolumn{1}{>{\centering\arraybackslash}m{2.5cm}}{\makecell[c]{\textbf{Key}\\\textbf{Features}}} \\
\Xhline{0.8pt}
\multirow{4}{*}{\makecell{ Water \\ Cherenkov }}
& Multi-ring topologies from relativistic charged particles and photons
& Secondary signatures from nuclear interactions and delayed decays
& Secondary nuclear de-excitation $\gamma$-ray emission  
& Large fiducial mass and long term stability \\
\hline
\multirow{5}{*}{\makecell{LArTPC}}
& 3-D imaging of charged particle final states
& Low momentum hadron reconstruction (e.g.  kaons)
& Partial kinematic reconstruction augmented by low energy hadrons and vertex activity
& High resolution imaging and particle identification capability\\
\hline
\multirow{5}{*}{\makecell{ Scintillators }}
& Total energy deposition from   charged particles
& Sensitivity to low energy and delayed decay products
& Prompt and delayed coincidence signatures associated with nuclear de-excitation
& Low energy threshold and excellent light yield \\
\hline
\multirow{4}{*}{\makecell{ Tracking \\ calorimeters }}
& Precision tracking and calorimetric reconstruction
& Identification of complex, multi-body topologies
& Partial reconstruction using observed charged particles
& Detailed  topology and kinematics reconstruction \\
\hline
\multirow{3}{*}{\makecell{ Radiochemical \\and geochemical }}
& Not applicable (no event level reconstruction)
& Not applicable (no event level reconstruction)
& Mode-independent sensitivity to rare decays
& Integrated sensitivity over long exposure times \\
\Xhline{1.2pt}
\end{tabular}
\caption{Complementarity of representative experimental approaches for $B$-violating processes.}
\label{tab:bexpcomp}
\end{table}

Liquid scintillator detectors, such as KamLAND~\cite{KamLAND:2005pen}, BOREXINO~\cite{Borexino:2003igu} and JUNO~\cite{JUNO:2025fpc}, operate by converting the energy deposited by charged particles into photons through  excitation and de-excitation processes, with a high scintillation light yield that enables low energy thresholds and efficient detection of low momentum and delayed signatures. These features are advantageous for $B$-violating processes where tagging of delayed decay products or nuclear de-excitation improves discrimination, including partially invisible decays accompanied by correlated low energy emission. Liquid scintillator detectors are advantageous to search for invisible nucleon decays. 
The experimental signature
arises from nuclear de-excitation, neutron emission or subsequent radioactive decays of the
residual nucleus.   

LArTPC detectors, such as ICARUS~\cite{Rubbia:2011ft}, provide three-dimensional imaging
of charged particle tracks together with calorimetric information derived from ionization charge. This can be leveraged for nucleon decay searches~\cite{Bueno:2007um}, enabling detailed reconstruction of event topology and particle identification based on track
morphology and energy losses, including direct sensitivity to low-momentum hadrons that lie below
the Cherenkov threshold in water. In particular, LArTPCs can readily identify kaons through their short
ionizing tracks and characteristic decay chains, well suited for nucleon decay
channels such as $p \to \bar{\nu} K^+$.
These capabilities  illustrate the complementary role of high-resolution tracking detectors. Subsequent generations of LArTPC detectors, such as DUNE~\cite{DUNE:2020lwj}, can extend this approach by providing complementary sensitivity through high resolution imaging and enhanced reconstruction of low momentum hadronic final states with increased exposure.

Additional approaches, including fine-grained tracking calorimeter experiments primarily based on iron such as NUSEX~\cite{Battistoni:1982vv}, KGF~\cite{Krishnaswamy:1985tho}, Soudan~\cite{Soudan-1:1986zji} and Frejus~\cite{Frejus:1990myz}, have set constraints on certain $B$-violating processes and signature classes. Further, radiochemical and geochemical methods, such as employed in early studies of Ref.~\cite{Flerov:1958zz,EvansJr:1977zuj,Fireman:1979xr}, provide qualitatively distinct probes by integrating exposure over significant geochemical timescales $\mathcal{O}$(Myr-Gyr). This enables mode-independent searches for baryon instability through accumulated daughter products, with interpretations of results limited by signal retention efficiencies and complex geochemical systematics. 
Complementary sensitivity for $B$ violation has also been achieved using  low background solid-state detectors, as demonstrated by experiments such as germanium-based GERDA~\cite{GERDA:2023uuw} and the Majorana Demonstrator~\cite{Majorana:2024rmx}, which are primarily targeting $L$ violation $0\nu\beta\beta$.

Tab.~\ref{tab:bexpcomp} summarizes complementarity between detector classes for $B$ violation searches. Across these detector technologies modern analyses increasingly exploit advanced multivariate and pattern recognition techniques to enhance signal-to-background discrimination and event classification, particularly in regimes where traditional selection-based approaches are limited by complex topologies or correlated observables.
Observation of $B$-violating signals in more than one detector technology with independent systematics would provide particularly compelling evidence for physics beyond SM. Null results across complementary approaches can yield robust constraints on broad classes of theoretical frameworks.

\subsection{Complementarity of physical processes}

Experimental searches for $B$ violation are intrinsically sensitive to extremely rare
processes. Results for the absence of signal excesses over expected
backgrounds are conventionally reported as lower limits on the partial lifetime $\tau/B$ for
a given decay channel, where $B$ denotes the branching fraction of the process under consideration.
For a representative decay mode, the partial lifetime limit is commonly expressed as
\begin{equation}
\frac{\tau}{B} = \frac{\Delta t   \varepsilon   N_{\text{nuc}}}{N_{\text{CL}}},
\label{eq:lifetime_limit}
\end{equation}
where $\Delta t$ is the experimental exposure in units of target mass multiplied by observation time,
$\varepsilon$ is the total signal detection efficiency, $N_{\text{nuc}}$ is the number of nucleons per
unit target mass, and $N_{\text{CL}}$ is the number of signal events allowed at a specified confidence
level according to the adopted statistical procedure. While this provides a useful
benchmark for comparing experimental reach, the physical interpretation depends on
the underlying process.

Beyond complementarity among detector technologies, sensitivity to $B$ violation is
fundamentally shaped by the diversity of physical processes under investigation. Different decay
modes can probe distinct underlying structures and principles including relevant effective operators, symmetries and UV completions. As a
result, the pattern of observed or excluded physical processes carries information that extends beyond
any single lifetime limit.

\begin{figure}[t]
\centering
\includegraphics[width=0.65\textwidth]{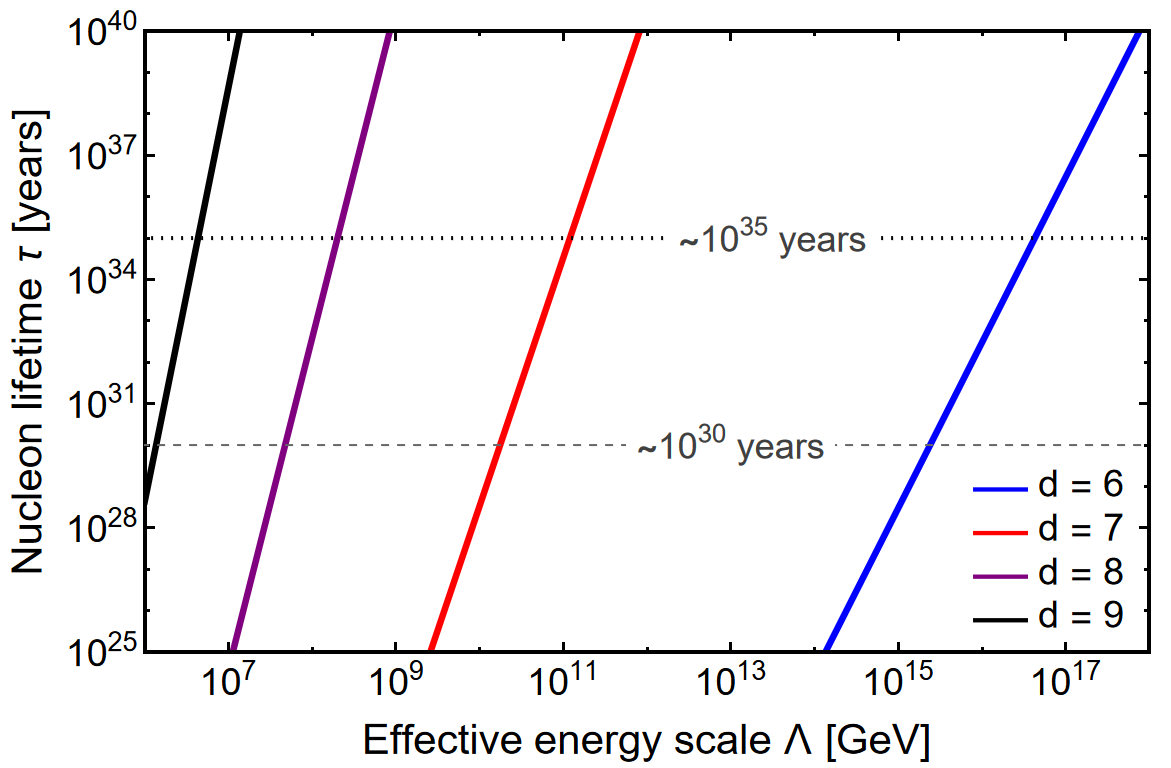}
\caption{Illustrative relationship between the effective scale $\Lambda$ of physics beyond SM and the nucleon lifetime
$\tau$ for $\Delta B = 1$ operators of different mass dimension $d$, assuming
order unity Wilson coefficients and dimensional analysis up to hadronic matrix elements and phase-space factors. Horizontal lines indicate representative benchmark lifetimes $\tau \sim 10^{30}$ years and $\tau \sim 10^{35}$ years. If $B$ violation is observed, a measured lifetime would map onto a corresponding scale
$\Lambda$ for each operator class, enabling interpretation of the underlying physics.
}
\label{fig:ndklifescale}
\end{figure}

From an EFT perspective $B$-violating interactions arise from
higher dimensional operators suppressed by a large mass scale $\Lambda$ associated with underlying microphysics. The relation between the
experimentally accessible nucleon decay lifetime and the corresponding scale $\Lambda$ depends sensitively on the
operator dimension. Fig.~\ref{fig:ndklifescale} illustrates this connection for $\Delta B = 1$
operators of different mass dimensions $d$, assuming for simplicity Wilson coefficients of order unity and dimensional
analysis following Eq.~\eqref{eq:tauScaling}. This clearly demonstrates that improvements in lifetime sensitivity translate into probes of qualitatively
different effective energy scales depending on the dominant governing operator structure. Thus, physics at different scales and associated theories play distinct roles.

Importantly, null search results in one class of decay modes do not generically exclude $B$ violation overall. Underlying structure selection rules, symmetries or accidental suppressions among others could eliminate the lowest dimension EFT operators responsible for characteristic proton decay channels while allowing higher dimensional
EFT operators or alternative channels to dominate. In such scenarios multi-body decays, channels involving
other particle combinations in final states  or processes with partially invisible final states could provide the dominant
experimental signatures. This motivates pursuing broad experimental programs that cover a wide range of decay
modes rather than focusing exclusively on a small subset of select benchmark channels.

Complementarity among physical processes is highlighted when comparing $\Delta B = 1$
and $\Delta B > 2$ observables. While $\Delta B =1 $  nucleon decay is typically sensitive to very high mass scales such as those that approach unification,
processes such as $\Delta B = 2$ oscillations of $n - \bar{n}$ or dinucleon decays arising from higher dimensional
effective operators that violate $B-L$ can probe qualitatively different symmetry structures and energy
regimes. Observation of $\Delta B > 1$ processes in the absence of single nucleon decays would therefore point
toward a distinctive origin of $B$ violation.

Taken together, the systematic study of multiple $B$-violating processes provides a
powerful diagnostic framework. The presence or absence of specific channels and their relative
sensitivities can encode rich information about the underlying theory, including the characteristic operator dimension, flavor structure  and symmetry content. Thus, complementarity among physical processes is not
just a technical consideration,  but a central element of the experimental strategy for unraveling the possible
origin of $B$ violation.

\subsection{Multi-nucleon decays  and neutron-antineutron oscillations ($\Delta B\geq 2$) }

Processes that violate $B$  with $\Delta B \geq 2$ can probe classes of physics scenarios that are qualitatively distinct from those associated with single nucleon decay. In contrast to $\Delta B = 1$ processes that are typically mediated by lower dimensional effective operators, $\Delta B \geq 2$ processes arise from higher dimensional operators and can involve symmetry structures beyond those tested by single nucleon decays. In a variety of motivated scenarios $\Delta B = 2$ transitions are often associated with $B-L$ violation, as  exemplified by $n - \bar n$. Hence, searches for $n - \bar n$ and multi-nucleon decays provide sensitivity to symmetry realizations and characteristic mass scales that could be distinct from single nucleon decay searches.

Among $\Delta B = 2$ processes, $n - \bar n$ oscillations are especially prominent and can appear in a variety of theoretical scenarios
and admit both free-particle (vacuum) and intranuclear experimental searches~(see e.g. Ref.~\cite{Phillips:2014fgb}), as originally proposed in early theoretical studies~\cite{Kuzmin:1970nx,Mohapatra:1980qe}. The associated $B - L$ structure can link such process to cosmological baryogenesis, and simultaneous observation of $\Delta B = 1$ proton decay and $\Delta B = 2$ $n - \bar{n}$ could be suggestive of underlying structure in which $\Delta L = 2$ Majorana neutrino masses arise~\cite{Babu:2014tra} and thus establishing complementarity.
Notably, $n - \bar{n}$ oscillations are associated in SMEFT with the leading $\Delta B = 2$ operator contribution appearing at dimension $d = 9$ as in Eq.~\eqref{eq:d9nnbar}, implying sensitivity to intermediate mass scales that are well below the typical unification scales. More so, $n - \bar n$ can occur as a coherent quantum oscillation, and admit both free-particle and intranuclear experimental searches with distinctive signatures and minimal model dependence.
Early studies also explored the possibility of hydrogen-antihydrogen ($H - \bar{H}$) oscillations as an atomic manifestation of $\Delta B = 2$~\cite{Feinberg:1978sd}. However, in practice, such transitions are highly suppressed. Consequently, $H - \bar{H}$ oscillations offer weaker sensitivity than $n - \bar{n}$ oscillations.

Experimentally, searches for $n - \bar n$ oscillations can be conducted either with free neutrons in vacuum or with neutrons bound in nuclei. For free neutrons, the process can be described by a simple two level effective Hamiltonian in the $\{|n\rangle, |\bar{n}\rangle\}$ basis, with off-diagonal elements parameterizing the $B$-violating transition. For free neutrons, the oscillation probability grows quadratically with observation time~\cite{Mohapatra:2009wp} 
\begin{equation}
P_{n\to\bar n}(t) \simeq \left( \frac{t}{\tau_{n\bar n}} \right)^2 ,
\label{eq:nnbartimefree}
\end{equation}
where $\tau_{n\bar n}$ is the oscillation time. Here, $\tau_{n\bar n} = 1/\delta m$, and the off-diagonal matrix element $\delta m$ parametrizes at the macroscopic nucleon level the $B$-violating microscopic physics that mediates the transition. Considering $d=9$ contributing operator in the context of SMEFT, one can estimate the matrix element $\delta m \sim \Lambda_{\rm QCD}^6/\Lambda^5$, where $\Lambda_{\rm QCD} \sim 200$~MeV
is the QCD scale accounting for the non-perturbative hadronization of quarks to $n$ and $\bar{n}$.  

Searches for free $n - \bar{n}$ oscillations typically exploit intense cold neutron sources and long flight paths, allowing to maximize the coherent observation time in Eq.~\eqref{eq:nnbartimefree}. 
If $\bar{n}$ contributions arise   during $n$ beam propagation they are revealed   by the annihilation on a target. This produces characteristic multi-meson final states, predominantly pions.
 Since even weak magnetic fields can lift the degeneracy between $n$ and $\bar{n}$ states, stringent magnetic shielding is necessitated. Reactor-based neutron facilities are particularly well suited to such searches, providing high fluxes of low energy neutrons and the infrastructure needed for extended vacuum beamlines. Stringent limits on free neutron oscillations have been  obtained at the Institut Laue-Langevin (ILL), corresponding to a lower bound on the oscillation time of order $\tau_{n \bar{n}}  \gtrsim 10^8$ s~\cite{Baldo-Ceolin:1994hzw}.
Experiments employing high intensity neutron sources, such as the European Spallation Source (ESS)~\cite{Santoro:2020nke} and other facilities, can enhance sensitivity to free $n - \bar{n}$ oscillations through  increased cold-neutron flux and longer coherent observation times.

In contrast, for bound neutrons  in nuclei the $n - \bar{n}$ oscillations are significantly modified by nuclear effects. The presence of the nuclear potential lifts the degeneracy between $n$ and $\bar{n}$ states, suppressing coherent oscillations. Instead, the experimental signatures arise  from the annihilation $\bar{n}$ with a neighboring nucleon, producing a multi-meson final state with total energy of $\sim 2$~GeV. The corresponding intranuclear transition rate is related to the free oscillation time via
\begin{equation}
\tau_{\text{nuc}}  = R_{\text{nuc}} \tau_{n\bar n}^2,
\end{equation}
where $R_{\text{nuc}}$ is a nucleus-dependent suppression factor that encodes nuclear structure and annihilation dynamics. Determination of $R_{\text{nuc}}$ relies on nuclear modeling~\cite{Friedman:2008es}, and introduces an unavoidable uncertainty when converting intranuclear lifetime limits into constraints on the free oscillation time.

Large underground detectors provide sensitivity to intranuclear $n - \bar n$ oscillations through their ability to identify the characteristic high-multiplicity hadronic final states. Water Cherenkov detectors exploit the complex signal topologies produced by multiple relativistic mesons, leading to stringent limits from Super-Kamiokande to  $\tau_{\rm nuc} \gtrsim 10^{32}$~years and beyond that can be converted to $\tau_{n \bar{n}} \gtrsim 10^{8}$~s~\cite{Super-Kamiokande:2020bov}. 
In a complementary approach tracking detectors with fine spatial resolution,  such as LArTPC~\cite{MicroBooNE:2023dci},  offer enhanced capability to reconstruct and discriminate individual signal tracks and vertices in complex $\bar{n}$ annihilation events.  

Related to $n-\bar{n}$ oscillations are searches for multi-nucleon decays, including dinucleon and trinucleon processes, which induce $\Delta B \geq 2$ and can arise from the same underlying effective operators. When multiple nucleons participate, the greater total available energy release can lead to distinct signatures and appearance of heavier 
particles such as $\tau$ leptons and $n p \to \tau^+\nu$ channels~\cite{Super-Kamiokande:2015pys} as well as higher momentum final states. 
Searches for dinucleon decay modes such as $pp, nn, np \to ({\rm mesons})$   have been performed in large underground detectors, yielding lower limits on partial lifetimes at the level of or exceeding $\mathcal{O}(10^{32})$~years~\cite{Super-Kamiokande:2014hie,Super-Kamiokande:2015jbb}. Trinucleon decay searches extend sensitivity to even higher dimensional operators but are experimentally more challenging. 
Across these searches, the dominant backgrounds arise from atmospheric neutrino interactions, especially producing multi-hadron final states. Background discrimination relies on sophisticated analyses. As with $\Delta B = 1$ searches, complementarity in detector technology enables sensitivity to a broader range of final states, backgrounds  and systematic uncertainties.

In addition to dedicated nucleon decay detectors, searches for multi-nucleon $B$ violation have been performed as ancillary analyses in experiments primarily designed to search for $0\nu\beta\beta$. These experiments employ very low backgrounds, high resolution detectors and long exposure times, enabling sensitivity to rare nuclear processes beyond their primary physics targets.
High purity germanium experiments such as GERDA and the Majorana Demonstrator, which search for $0\nu\beta\beta$ decay in $^{76}$Ge, have conducted searches for trinucleon decays inducing $\Delta B = 3$  by exploiting distinctive nuclear signatures associated with the disappearance of three nucleons and the subsequent radioactive decay of unstable daughter nuclei. They have established lower limits on decay lifetimes exceeding $\mathcal{O}(10^{25})$ years, providing stringent constraints on several $\Delta B \ge 3$ channels to date~\cite{GERDA:2023uuw,Majorana:2024rmx}.

While $\Delta B \ge 3$ processes probe higher dimensional operators and are typically subdominant to single  or dinucleon decays, such searches are theoretically motivated in scenarios where lower order $B$-violating operators are forbidden or suppressed. These results highlight an important experimental synergy between $B$ violation and $L$ violation programs, which benefit from similar requirements on background suppression, detector stability  and precision nuclear spectroscopy.

\subsection{Inclusive and model-independent searches}

Beyond channel specific searches for nucleon decays into fully visible final states, a powerful and complementary experimental strategy is provided by inclusive and model-independent searches for $B$ violation~\cite{Heeck:2019kgr}. These searches are formulated to be largely agnostic to the detailed particle content and enable the simultaneous probing of broad classes of modes, including non-canonical channels involving additional
possible light, weakly interacting or invisible particles \cite{Fridell:2023tpb}.

At the effective level, for single nucleon $N$ decay, such processes could be schematically represented as
\begin{equation}
N  \to  X_{\rm vis} + X_{\rm inv} ,
\end{equation}
where $X_{\rm vis}$ denotes visible final states and $X_{\rm inv}$ denotes ``invisible'' final states that are weakly interacting, kinematically
suppressed or otherwise undetectable within a given experimental configuration. 
The notion of an invisible final state encompasses a broad range of physical
possibilities.
Invisible components may arise from particles with energies below experimental detection
thresholds, from low energy or absorbed secondary contributions produced inside a nucleus or from
long lived particles that decay outside the active detector volume.
Particles produced invisibly at the decay vertex could later undergo
secondary interactions or decay into visible states, leading to delayed or spatially
separated signatures. Inclusive searches do not need to be restricted to fully invisible final states. 

A prominent class of signatures relevant for inclusive nucleon decay searches is associated with nucleon decay into a single visible charged lepton accompanied by additional neutral particles, such as neutrinos. This results in multi-body final states with a continuous charged lepton energy spectrum rather than a mono-energetic line characteristic of two-body decays, as exemplified by searches for channels such as $p \to e^+\nu\nu$~\cite{Super-Kamiokande:2014pqx}. Consequently, the signal appears as a spectral distortion or excess relative to backgrounds. Spectral searches of this type retain sensitivity to $B$ violation mediated by higher dimensional operators and multi-body dynamics, even when only one visible final state particle is experimentally accessible~\cite{Super-Kamiokande:2015pys}. Inclusive search strategies can also be formulated for channels involving other visible final state particles including mesons or photons, while retaining sensitivity to broad classes of $B$-violating processes.

More generally, inclusive searches enable simultaneous sensitivity to a variety of processes without imposing restrictions on specific final state assumptions. 
They can be defined by requiring the presence of a limited subset of
visible particles or event features, while remaining agnostic to the rest of the
final state composition.
Such
analyses enable  probing limits on the partial lifetime $\tau_N$ simultaneously for whole families of decay channels
\begin{equation}
\tau_N^{-1}  =  \sum_i \Gamma(N \to X_i) ,
\end{equation}
where the sum index $i$ accounts for all processes contributing to selected   signatures.
This approach also provides a robust probe of $B$ violation in scenarios where the branching
fractions among individual processes is not specified or known.   

Besides enabling broad coverage of physical processes with SM final states, inclusive searches enable efficient probes of non-canonical nucleon decay channels with additional light states beyond SM. 
In the simplest limiting case, the additional state may be effectively invisible and weakly interacting, leading to signatures of the form $p \to l^+X$, where $l$ is a lepton and $X$ carries away 
missing energy without further observable activity. Such searches make minimal assumptions about the microscopic nature of the invisible state and are sensitive to scenarios in which $B$ violation couples to hidden or sterile parts of the theory. These signatures of charged lepton with missing energy provide a controlled extension of conventional nucleon decay searches covering a variety of models~\cite{Super-Kamiokande:2015pys}. Non-canonical nucleon decays can arise in various contexts, including   neutron decays with dark sector particles~\cite{Davoudiasl:2014gfa,Fornal:2018eol} and nucleon decays involving right handed neutrinos~\cite{Helo:2018bgb}. Consequently, $B$ violation need not appear through
conventional visible final states and may evade exclusive searches. Non-canonical nucleon decays significantly broaden the possibilities of which physical processes can occur, and can include additional dark photon (vector), axion-like particle (pseudo-scalar), sterile neutrino (fermion), Majoron (scalar) and other states, as has been   systematically explored in a framework encompassing a wide range of new physics scenarios in Ref.~\cite{Fridell:2023tpb} that include light new particles with masses $\lesssim \mathcal{O}$(GeV).
In this approach, missing energy, soft visible spectra and multi-body final states
emerge as generic features, providing strong motivation for inclusive and
model-independent search strategies.

Fully invisible decay modes constitute limiting cases that motivate general model-independent searches. This is exemplified by 
processes such as $n  \to  \nu \nu \nu$ that can arise in models of extra dimensions~\cite{Mohapatra:2002ug}, and here  $\nu$ and $\bar{\nu}$ are not being distinguished. More generally, a large class  of fully invisible decays is possible in the context of new weakly interacting particles beyond SM~\cite{Fridell:2023tpb}.
Searches for invisible modes do not rely on specific models, and are typically carried out by considering nuclear response to the sudden loss of a bound nucleon. The daughter nucleus is left in an excited state, de-exciting via characteristic MeV-scale gamma rays and occasionally other particles~\cite{Ejiri:1993rh,Kamyshkov:2002wp}. These de-excitation gamma searches cover all invisible modes, whether SM neutrinos, beyond SM particles as well as processes outside of experimental detection regimes for a given detector. Another approach to enhance visibility of nucleon decays is to consider their external decay outside of the experiment that would result  in secondary signatures inside the detector. This is exemplified by analyzing neutrino flux in the detector produced from   nucleon decays to neutrinos inside the Earth outside of experiment~\cite{Learned:1979gp}. Similar strategy can also be considered for particles beyond SM decaying outside of detector, such as sterile neutrinos~\cite{Heeck:2025uwh}.

The sensitivity of different detector technologies to these inclusive and model-independent search signatures can vary
substantially. Detectors with lower energy thresholds and
improved gamma sensitivity, such as liquid scintillator experiments or heavy-water detectors could offer
enhanced capabilities for searches involving low energy signatures. In particular, the ability to detect low-energy gamma
cascades and to discriminate them from natural radioactive backgrounds makes such experiments
especially well suited for invisible or near-invisible decay modes. Inclusive and model-independent searches therefore play a key role in extending experimental
coverage beyond canonical decay modes. They provide sensitivity to a wide range of new physics
scenarios, including those involving additional light states, hidden sectors, or exotic spacetime
structure, and ensure that experimental programs remain sensitive to unexpected manifestations of
baryon number violation even in the absence of well-defined benchmark models.

\subsection{Externally induced and catalyzed processes}

In addition to spontaneous nucleon decays arising from intrinsic instabilities of baryons,
$B$ violation can also occur through processes that are  induced or catalyzed externally
by particles or macroscopic objects traversing a detector or an external environment.
In such scenarios, $B$-violating interactions do not correspond to an intrinsic decay
property of isolated nucleons, but are instead triggered by the presence of an incident agent
carrying nontrivial topological, quantum or baryonic structure. Thus, related experimental
searches should be interpreted in terms of incident fluxes and event rates rather than nucleon lifetimes, although
observable final states could resemble those of spontaneous nucleon decay.

A broad class of theoretical frameworks predicts such externally induced $B$-violating
processes. Representative examples include macroscopic magnetic monopoles, macroscopic non-topological solitons such
as $Q$-balls as well as microscopic dark sector particles carrying $B$ charge or mediating $B$-violating
interactions. Despite their diverse microscopic origins, these scenarios can lead to $B$ violation localized along the trajectories of incident objects, giving rise to  signal rates proportional to the incident flux rather than solely to detector exposure. In specific interpretations, assumptions about the origin or population of the incident objects
may be used to relate their flux to external constraints.

In the presence
of monopoles that could originate from GUTs~\cite{tHooft:1974kcl,Polyakov:1974ek}, and in the early Universe~\cite{Preskill:1979zi}, $B$ violation may be catalyzed through 
the Callan-Rubakov mechanism~\cite{Rubakov:1981rg,Callan:1982ah}. A monopole traversing matter could  induce nucleon
decays along its trajectory. The expected event rate in a detector can be written schematically as
\begin{equation} 
\label{eq:macroflux}
R  =  \Phi_M   \sigma_{\rm cat}   N_T ,
\end{equation}
where $\Phi_M$ is the monopole flux, $\sigma_{\rm cat}$ is an effective catalysis cross-section 
and $N_T$ is the number of target nucleons. 
Experiments have carried out dedicated searches for related signatures. Complementary to direct searches for catalyzed nucleon decays inside detectors, experiments can also look for indirect signatures of nucleon decays catalyzed elsewhere. This is illustrated by Super-Kamiokande searches for neutrinos produced by monopoles catalyzing nucleon decays inside the Sun~\cite{Super-Kamiokande:2012tld}. 
Related phenomena can also arise for non-topological solitons, such as $Q$-balls~\cite{Coleman:1985ki},
which are macroscopic field configurations carrying conserved global charges such as $B$ and $L$.  Experimental searches have investigated signatures of baryonic Q-ball interactions as they traverse ordinary matter and catalyze nucleon decays, as searched by Super-Kamiokande~\cite{Super-Kamiokande:2006sdq}.

Dark sector scenarios can provide realizations of externally induced $B$ violation by microscopic interactions.
In models such as hylogenesis~\cite{Davoudiasl:2010am} or related mesogenesis~\cite{Berger:2023ccd},
the dark sector particles can carry conserved $B$ opposite to that of visible matter, linking the
cosmological baryon asymmetry to the dark matter abundance. Interactions between such
dark matter particles and ordinary nucleons can then mediate $B$-violating transitions. Experimentally, these processes
can induce nucleon decay-like final state signatures with modified kinematics.  
Although the visible final states of externally induced $B$ violation can
mimic those of non-canonical or inclusive spontaneous nucleon decay searches, such as decays
involving light new particles or modified kinematics, the underlying physical interpretation is
distinct. This motivates
separate experimental strategies and interpretations.

Such searches broadly extend the scope of experimental tests of $B$
conservation beyond conventional spontaneous decay processes.

\subsection{Cosmological and astrophysical constraints and probes}
\label{sec:astro_constraints}

Complementary to laboratory-based experiments, $B$ violation can also be constrained and in some cases tested through astrophysical and cosmological observations. These environments
provide access to extreme conditions such as high densities, large energies  and long timescales, challenging to access on Earth. This enables probing $B$ conservation in complementary
regimes. While such analyses and constraints are typically indirect and often depend on additional assumptions and modeling, they play a relevant role in delineating the viable parameter space of $B$-violating theories
and in establishing consistency conditions   complementary to controlled laboratory searches.

The observed matter-antimatter asymmetry of the Universe restricts
$B$-violating interactions in the early Universe. Any $B$-violating
processes that remain in thermal equilibrium after the generation of the baryon asymmetry
can erase earlier existing asymmetry~\cite{Kolb:1979qa,Kuzmin:1985mm}, a phenomenon known as washout.
This constrains the strength and temperature dependence of possibly active
$B$-violating processes and motivates scenarios in which $B$ violation
occurs out of equilibrium or is accompanied by $B-L$ violation, taking into account electroweak sphaleron effects.  Primordial Big Bang nucleosynthesis also provides indirect tests of $B$-violating processes at early times~\cite{Sarkar:1995dd}, although generally weaker than those arising from baryon washout considerations.
These cosmological consistency conditions play an important role for viable frameworks of $B$ violation.

Astrophysical objects can provide indirect but prominent probes of $B$ violation by testing the stability of dense, bound nuclear matter over astrophysical
timescales. In neutron stars and white dwarfs nucleons reside in dense degenerate matter where in-medium potentials and Pauli blocking qualitatively modify their kinematics relative to free particles, strongly suppressing free neutron beta decay.  $B$-violating processes can
be affected in nontrivial ways by in-medium effects. For example, $\Delta B = 2$ transitions like $n - \bar{n}$ in dense matter  could lead to efficient annihilation,
heating or mass loss incompatible with the observed stability of neutron stars.
Such considerations have been used to derive qualitative constraints on $B$-violating operators~\cite{Berryman:2022zic}. 
Related arguments have also been considered to constrain $B$-violating processes with dark sector final states~\cite{Baym:2018ljz,McKeen:2018xwc}.
Interpreted conservatively, and depending on underlying assumptions, these arguments correspond to effective $B$-violating lifetimes exceeding
$\sim 10^{31}-10^{33}$~years. Constraints on $\Delta B = 1$ processes from
compact objects are generally weaker and more model-dependent. Hence, astrophysical stability arguments
provide complementary information.

While cosmological and astrophysical probes are indirect and often rely on additional assumptions and model dependencies compared with laboratory searches with various controlled systematics, together they form a multi-faceted strategy for comprehensively exploring   possible $B$ violation.

\section{Experimental Tests of Lepton Number Conservation}

\subsection{Neutrino oscillations and lepton flavor violation $(\Delta L = 0)$}

In the broader landscape of experimental tests of $L$ and lepton flavor, neutrino oscillations play a
foundational role by demonstrating the non-conservation of individual lepton flavor numbers.
Consequently, they establish lepton flavor violation among neutrinos while conserving total $L$, with $\Delta L = 0$. This provides  a key point of reference for the interpretation of searches for CLFV and $\Delta L \neq 0$ processes.

Neutrinos play a special role among known fundamental fermions, since they do not carry electric or color charge and within SM interact only through weak interactions and gravity.  In the original formulation of SM neutrinos were assumed to be   massless 
and individual lepton flavor numbers  $L_e$, $L_\mu$ and $L_\tau$  were independently conserved.  
However, it has been established experimentally that neutrinos undergo
flavor oscillations~\cite{Super-Kamiokande:1998kpq}, implying that they possess non-zero masses and can mix among flavors
\cite{Pontecorvo:1957qd,Maki:1962mu}.

Neutrino oscillations arise  from a mismatch between the weak interaction
(i.e. flavor) eigenstates $(\nu_e,\nu_\mu,\nu_\tau)$ and the mass eigenstates $(\nu_1,\nu_2,\nu_3)$.
As neutrinos propagate, quantum interference between different mass eigenstates causes their
flavor composition to evolve with distance and energy. Since the conversion probability $P(\nu_\alpha \to \nu_\beta)$ between neutrino species $\alpha$ and $\beta$ depends on their non-zero   mass-squared differences $\Delta m^2$, the observation of such flavor transitions requires that neutrinos are massive.
Originally proposed by Pontecorvo in analogy with $K^0-\bar{K}^0$ neutral kaon mixing  
\cite{Pontecorvo:1957qd} and subsequently confirmed through a wide range of experiments,
including solar neutrino measurements \cite{Davis:1968cp,SNO:2002tuh}, atmospheric neutrino observations
\cite{Super-Kamiokande:1998kpq} as well as reactor~\cite{KamLAND:2004mhv} and accelerator~\cite{K2K:2002icj} studies (see~Ref.~\cite{ParticleDataGroup:2024cfk} for a comprehensive overview).

Neutrino masses, introducing scales far below those
 of charged leptons and quarks, necessitate an extension of the renormalizable SM interactions. Notably, neutrino oscillations do not violate $L$ and do not determine whether neutrinos are Dirac or Majorana
particles~\cite{Majorana:1937vz}.
Since neutrinos are electrically neutral, they admit the possibility of Majorana mass terms such that neutrinos and antineutrinos are identical.
If neutrinos are Majorana particles then $L$  is not an exact symmetry and processes with
$\Delta L = 2$ may occur.
A particularly sensitive probe of this possibility is $0\nu\beta\beta$
\cite{deGouvea:2013zba}.
Determining whether neutrinos are Dirac or Majorana fermions thus also tests if $L$ is a fundamental symmetry of Nature. This can have far reaching implications,  
including possible connections to $B$ violation and the origin of the cosmic
matter-antimatter asymmetry~\cite{Fukugita:1986hr}.

Although neutrino oscillations establish that neutrino masses are non-zero, they are not sensitive to the absolute mass scale. Direct kinematic probes of neutrino masses, such as precision measurements of the tritium  $\beta$-decay close  to its kinematic endpoint, provide a complementary and model-independent approach. This is exemplified by the KATRIN~\cite{KATRIN:2024cdt} experiment, which restricts effective electron neutrino mass $m_{\nu}$ to sub-eV regimes in laboratory setting. Notably, such measurements are independent of the neutrino's possible Dirac or Majorana nature and also do not rely on cosmological assumptions. Future techniques, including approaches based on atomic and molecular spectroscopy, may further extend the sensitivity of direct searches into the sub-eV regime.

\subsection{Charged-lepton flavor violation $(\Delta L = 0)$}

The discovery of neutrino oscillations established that individual lepton flavor numbers
$L_e$, $L_\mu$, and $L_\tau$ are not conserved in Nature.
All experimentally confirmed flavor changing phenomena to date occur in the neutrino sector
and conserve $L$.
Searches for CLFV extend this to charged leptons~\cite{Calibbi:2017uvl},
probing whether transitions such as $\mu \to e\gamma$, $\mu \to 3e$ or coherent
$\mu-e$ conversion in nuclei can occur. The canonical CLFV benchmark processes conserve total $L$, with $\Delta L = 0$.  These processes are distinct from interactions violating $L$ with $\Delta L \neq 0$,
such as $0\nu\beta\beta$, which are complementary and probe fundamentally different underlying
structures and symmetries.

In the minimal SM extended to include non-vanishing neutrino masses  CLFV processes are allowed through loop diagrams involving  light active neutrinos and $W$ bosons \cite{Cheng:1980tp}. However, the amplitudes are strongly suppressed by the small neutrino masses and the Glashow-Iliopoulos-Maiani (GIM) mechanism \cite{Glashow:1970gm}. In particular, the branching fraction for $\mu \to e\gamma$ scales as ${\rm Br}(\mu \to e\gamma) \propto (\Delta m_\nu^2/M_W^2)^2$ resulting in ${\rm Br}(\mu \to e\gamma) < 10^{-54}$ \cite{deGouvea:2013zba}. Such rates are significantly below   typical experimental sensitivities. Consequently, observable CLFV signals would imply new particles or interactions beyond the SM.
Historically, muon properties and decays have played a prominent role in revealing fundamental scales of particle physics \cite{Kuno:1999jp}. Muon lifetime measurements enable among the most precise determinations of Fermi constant $G_F$~\cite{MuLan:2012sih}, and CLFV decays enable probing new potential scales of physics beyond SM and associated flavor structure. 

CLFV searches should be distinguished from tests of lepton universality~\cite{Kuno:1999jp}. Universality 
probes test whether the coupling of leptons to gauge bosons is independent of flavor. In terms of flavor space, CLFV searches probe off-diagonal flavor interactions while universality searches probe diagonal couplings among lepton generations. These tests are complementary, and physics beyond SM may preserve lepton universality while violating lepton flavor conservation or vice versa. 

Many  motivated extensions of SM predict CLFV
at experimentally accessible levels (see Ref.~\cite{Lindner:2016bgg} for a comprehensive overview).
Some representative examples include models with heavy right handed neutrinos and seesaw mechanisms~\cite{Minkowski:1977sc,Gell-Mann:1979vob,
Yanagida:1979as}, GUTs~\cite{Georgi:1974sy}, SUSY theories~\cite{Hall:1985dx,Borzumati:1986qx}, leptoquarks~\cite{Davidson:1993qk} extended Higgs sectors~\cite{Cheng:1987rs}. At energies below the scale of new physics beyond SM, CLFV effects can be described with EFT framework by dimension-six operators~\cite{Grzadkowski:2010es} including dipole,
four lepton  and lepton-quark contact interactions~\cite{Buchmuller:1985jz}.
Different physical processes probe distinct operators and underlying organizing structures. 
Radiative decays such as $\mu \to e\gamma$ are primarily sensitive to dipole operators,
three-body decays to four lepton contact interactions 
and $\mu-e$ conversion in nuclei to lepton-quark operators.

Stringent constraints on CLFV arise from searches related to muons, which can benefit from
intense sources, precise  kinematics and significant background control~\cite{Kuno:1999jp}.
Unlike quark flavor-changing processes, CLFV searches typically  have significantly reduced backgrounds from SM contributions and thus enable sensitive tests of flavor symmetries. 
A long-serving benchmark channel  is $\mu^+ \to e^+\gamma$, with MEG stopping muon experiment constraining branching ratios at the level of ${\rm Br}(\mu^+ \to e^+\gamma) < \mathcal{O}(10^{-13})$~\cite{MEGII:2025gzr}.
Complementary searches target processes such as $\mu^+ \to e^+ e^+ e^-$ that have been stringently constrained to levels Br$< \mathcal{O}(10^{-12})$ by the SINDRUM experiment~\cite{SINDRUM:1987nra}, and can be probed with experimental approaches employing high intensity muon beams and precision tracking as exemplified by Mu3e~\cite{Berger:2014vba}.
Coherent $\mu-e$ conversion in the field of a nucleus  $\mu^- + (A,Z) \to e^- + (A,Z)$ 
probes lepton-quark contact interactions and depending on the target has been constrained to levels ${\rm Br} < \mathcal{O}(10^{-13})$ by the SINDRUM II experiment~\cite{SINDRUMII:2006dvw}, and can be further tested by intense pulsed muon beams and optimized background rejection such as in Mu2e~\cite{Mu2e:2022ggl} and COMET~\cite{COMET:2018auw} experiments.
Searches for CLFV in tau decays probe higher lepton masses and scenarios in which physics
beyond SM preferentially interacts with the third generation, with
processes such as $\tau \to \ell\gamma$   constrained
at the level of ${\rm Br} < \mathcal{O}(10^{-8})$~\cite{ParticleDataGroup:2024cfk}. $B$-factory experiments and high luminosity $e^+e^-$ collider facilities, such as Belle II~\cite{Belle-II:2018jsg} and BarBar and Belle~\cite{HeavyFlavorAveragingGroupHFLAV:2024ctg}, can sensitively probe such $\tau$ decay processes.

\subsection{Neutrinoless double beta decay ($\Delta L=2$)}

The observations of neutrino oscillations establish that neutrinos are not massless and that individual lepton flavor numbers are not conserved.
However, oscillation phenomena are insensitive to the absolute neutrino mass scale and cannot determine whether neutrinos are Dirac or Majorana fermions.
Determining the fundamental nature of neutrino masses is a central open question. As discussed, a prominent role in addressing this question as well as in probing $L$ violation is played by $0\nu\beta\beta$ decays~(for an overview, see Ref.~\cite{Avignone:2007fu,DellOro:2016tmg}), detection of which can lead to far reaching implications in particle physics and cosmology.

A direct  and sensitive experimental probe of $L$ violation with $\Delta L = 2$ is provided by $0\nu\beta\beta$ decay~\cite{Furry:1939qr}.
In this nuclear transition, depicted in Fig.~\ref{fig:processes}, one expects~\cite{Primakoff:1959chj}
\begin{equation} \label{eq:0nubbnu}
(Z,A) \rightarrow (Z+2,A) + 2e^- ,
\end{equation}
where in a nucleus with mass number $A$ and atomic number $Z$ two neutrons are converted into two protons with the emission of two electrons and no neutrinos.
Observation of this process would establish that $L$ is not an exact symmetry of Nature.
By contrast, the SM allows two neutrino double beta decay with ($2\nu\beta\beta$)  
\begin{equation}
(Z,A) \rightarrow (Z+2,A) + 2e^- + 2\bar{\nu}_e ,
\end{equation}
that conserves $L$ and has been observed in multiple isotopes.
Originally proposed in 1935 by Goeppert-Mayer~\cite{Goeppert-Mayer:1935uil}, measurements of $2\nu\beta\beta$ provide benchmarks for nuclear structure calculations~\cite{Elliott:1987kp,Barabash:2015eza}.

If neutrinos have Majorana mass terms, with physical mass eigenstates satisfying
$\nu = \nu^c$, then $L$ is violated and $0\nu\beta\beta$ decay becomes possible.
In the minimal realization the decay is mediated by the exchange of light active Majorana neutrinos that participate in the SM weak interactions.
Consequently, the decay amplitude is proportional to the effective Majorana mass
\begin{equation}
|m_{\beta\beta}| = \left| \sum_i U_{ei}^2  m_i \right|,
\end{equation}
where $U_{ei}$ are elements of the Pontecorvo-Maki-Nakagawa-Sakata (PMNS) matrix~\cite{Pontecorvo:1957qd,Maki:1962mu} and $m_i$ are the neutrino mass eigenvalues.
The general effective description for the $0\nu\beta\beta$  inverse half life introduced in Eq.~\eqref{eq:0nubb_rate} can then be written as  
\begin{equation}
\left[T_{1/2}^{0\nu}\right]^{-1}
= G^{0\nu}  |M^{0\nu}|^2   \frac{|m_{\beta\beta}|^2}{m_e^2},
\end{equation}
where $G^{0\nu}$ is a phase-space factor~\cite{Kotila:2012zza}, $M^{0\nu}$ is the nuclear matrix element and $m_e$ is electron mass.
In EFT this corresponds to the long-range light-neutrino exchange mechanism, which can be realized with $d = 5$ Weinberg operator in Eq.~\eqref{eq:weinbergop}.

The $0\nu\beta\beta$ decay is an especially powerful probe of $L$ violation. In contrast to $2\nu\beta\beta$ decay expected in SM, in $0\nu\beta\beta$ decay the two emitted electrons carry the whole emission energy. Hence, the main
experimental signature of $0\nu\beta\beta$ decay is a peak-like feature in the summed electron energy spectrum that is broadened by experimental resolution. The sensitivity arises from both theoretical and experimental considerations.
The leading $\Delta L = 2$ interaction in SMEFT is the $d = 5$ Weinberg operator that has the lowest mass dimension and therefore the weakest parametric EFT suppression among $\Delta L  = 2$ operators. 
However, this does not imply that such contributions necessarily dominate the $0\nu\beta\beta$ decay rate in all UV completions. In various scenarios short-range contributions associated with higher dimensional operators can provide the leading effects~\cite{Deppisch:2012nb}.
This enables sensitivity to high  effective scales $\Lambda$ of physics beyond SM compared to higher dimensional interactions.  Additionally, $0\nu\beta\beta$ benefits from nuclear coherence and from the possibility of accumulating large exposures with  low backgrounds in stable isotopic targets.  This yields sensitivity to $L$ violation scales that are typically challenging to access.

$L$ violation can arise from the spontaneous breaking of a global $L$ symmetry.
Consequently, the associated Majoron (pseudo-)Nambu-Goldstone boson $J$ can be emitted in double beta decays~\cite{Chikashige:1980ui,Gelmini:1980re,Georgi:1981pg}
\begin{equation}
(Z,A) \rightarrow (Z+2,A) + 2e^- + J ,
\end{equation}
which results in continuous electron energy spectrum that is distinct from that of standard $0\nu\beta\beta$ of Eq.~\eqref{eq:0nubbnu}.
Searches for Majoron emitting channels therefore probe qualitatively different realizations of $L$ violation and provide complementary information on the underlying symmetry structure.
 
Although light Majorana neutrino exchange provides a minimal benchmark, $0\nu\beta\beta$ decay can also be induced by a broad class of $L$-violating interactions arising from physics beyond SM. These include the possible exchanges of heavy Majorana neutrinos, SUSY particles, leptoquarks, extended Higgs sector fields and other mechanisms~\cite{Doi:1985dx,Haxton:1984ggj,Mohapatra:1986su,Mohapatra:1980yp,Hirsch:1996ye}.  At low energies such contributions can be described  by higher dimensional EFT operators~\cite{Pas:1999fc,Helo:2016vsi}.
Earlier EFT approaches classified $0\nu\beta\beta$ contributions into the light Majorana neutrino mass mechanism that is proportional to effective mass $m_{\beta\beta}$, long-range contributions not directly proportional to $m_{\beta\beta}$ and short-range contact-like contributions. Modern formulations employ chiral perturbation theory, which automatically incorporates the pion contributions~\cite{Cirigliano:2017djv,Cirigliano:2018yza}.

An observation of $0\nu\beta\beta$ decay implies the existence of Majorana masses for active neutrinos in any local  Lorentz invariant quantum field theory and independent of possible microphysics processes mediating it.
This is consequence of the Schechter-Valle ``black-box'' theorem~\cite{Schechter:1981bd}, which ensures that $\Delta L = 2$ interactions induce a Majorana mass term through radiative corrections.
However, the resulting induced Majorana mass could be exceedingly small and need not account for the observed neutrino masses~\cite{Duerr:2011zd}.
Consequently, while a discovery of $0\nu\beta\beta$ would establish the Majorana nature of neutrinos it would not by itself identify the dominant microscopic mechanism responsible for the decay.

\begin{figure}[t]
\centering
\includegraphics[width=0.65\linewidth]{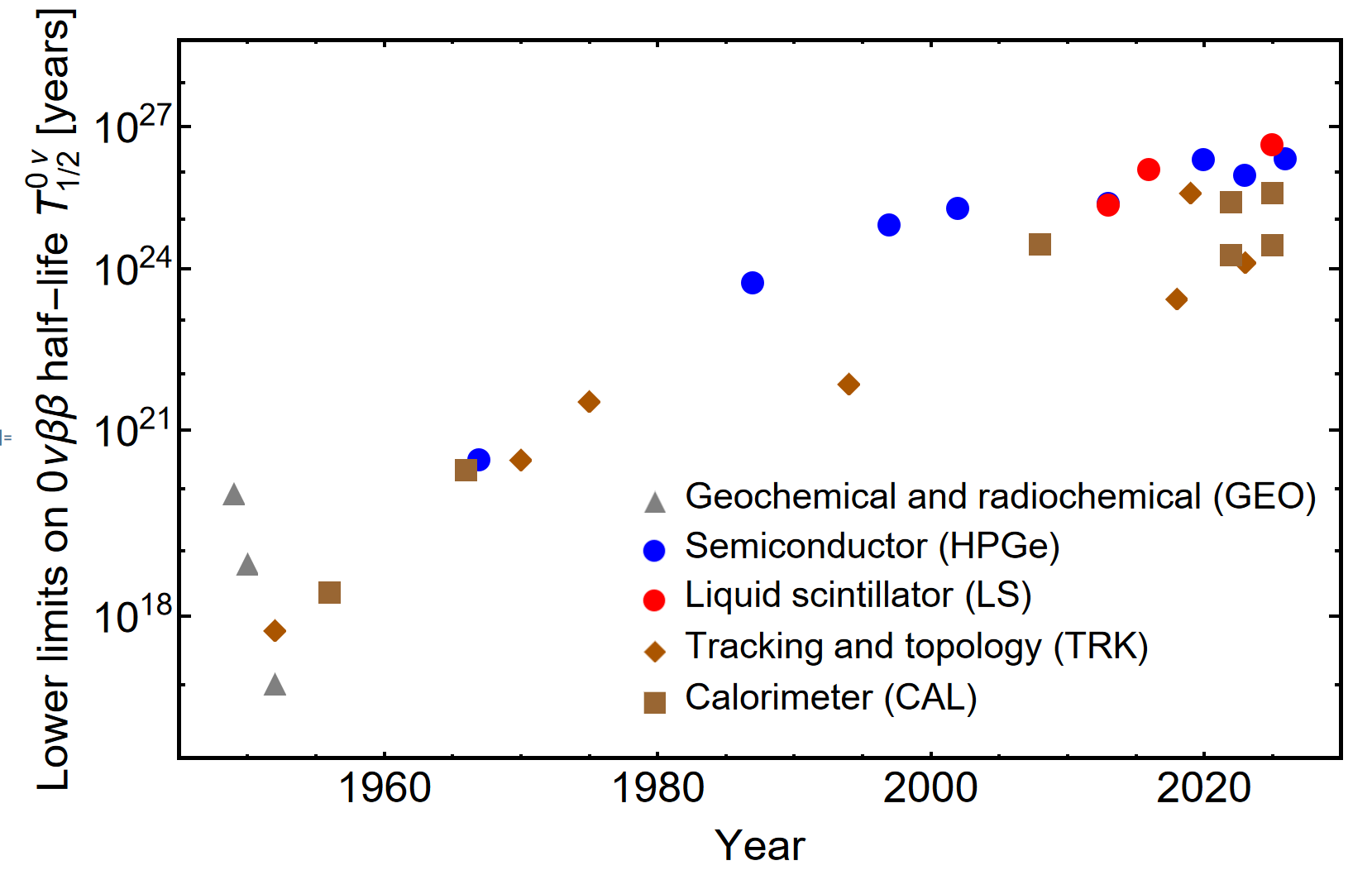} 
\caption{Historical evolution of experimental lower limits on the neutrinoless double beta decay ($0\nu\beta\beta$) half-life 
$T_{1/2}^{0\nu}$.
Representative results illustrate the development of distinct experimental
approaches rather than an exhaustive compilation.
Detection principles are denoted as follows: early radiochemical and geochemical constraints (GEO, including
$^{130}$Te~\cite{Inghram:1949qu}, $^{124}$Sn~\cite{Fireman:1952qt},
$^{238}$U~\cite{Levine:1950qw},  $^{150}$Nd~\cite{Cowan:1956pz}),
high purity germanium semiconductor detectors (HPGe, primarily using $^{76}$Ge, including early
measurements~\cite{Fiorini:1967in,Caldwell:1987gc} as well as Heidelberg-Moscow~\cite{Gunther:1997ai}, IGEX~\cite{IGEX:2002bce}, GERDA~\cite{GERDA:2013vls,GERDA:2020xhi},
Majorana Demonstrator~\cite{Majorana:2022udl}, LEGEND-200~\cite{LEGEND:2025jwu} experiments),
isotope loaded liquid scintillator detectors (LS, notably $^{136}$Xe in KamLAND-Zen~\cite{KamLAND-Zen:2012mmx,KamLAND-Zen:2016pfg,KamLAND-Zen:2024eml} experiment),
tracking detectors with event topology reconstruction (TRK, including early cloud chamber and streamer chamber
experiments with $^{110}$Pd~\cite{Winter:1952zz} and $^{82}$Se~\cite{Cleveland:1975zz}, as well as
NEMO-2~\cite{NEMO:1994sst}, NEMO-3~\cite{Arnold:2018tmo},
EXO-200~\cite{EXO-200:2019rkq}, and NEXT~\cite{NEXT:2023daz} experiments),
and calorimetric detectors (CAL, including crystal scintillator experiments such as early $^{48}$Ca measurements~\cite{derMateosian:1966qz,Bardin:1970vvi}, 
and cryogenic calorimeters (bolometers), including $^{100}$Mo-based CUPID~\cite{Augier:2022znx} and AMoRE~\cite{AMoRE:2024loj}, 
as well as Te-based CUORICINO~\cite{CUORICINO:2008jjc} and CUORE~\cite{CUORE:2021mvw,CUORE:2024ikf}).
Results from different isotopes are shown together, but interpretation in terms of
particle parameters depends on specific phase-space factors and nuclear matrix elements. A confirmed detection signal would manifest as a finite measured half-life, and the historical record contextualizes sensitivity evolution. }
\label{fig:0nubbhist}
\end{figure}

Fig.~\ref{fig:0nubbhist} summarizes the historical evolution of experimental lower limits on the
$0\nu\beta\beta$ half-life. Representative experimental approaches rather than an
exhaustive record of the most stringent limits achieved at each time are shown.
Detection principles are grouped into radiochemical and geochemical methods (GEO)~\cite{Inghram:1949qu,Fireman:1952qt,Levine:1950qw,Cowan:1956pz},
calorimetric source and detector experiments (CAL)~\cite{derMateosian:1966qz,Bardin:1970vvi,CUORICINO:2008jjc,CUORE:2021mvw},
high purity germanium semiconductor detectors (HPGe)~\cite{Fiorini:1967in,Caldwell:1987gc,Gunther:1997ai,IGEX:2002bce,GERDA:2013vls,GERDA:2020xhi},
liquid scintillator detectors with isotopes (LS)~\cite{KamLAND-Zen:2012mmx,KamLAND-Zen:2016pfg,KamLAND-Zen:2022tow} 
as well as tracking detectors with event topology reconstruction (TRK)~\cite{Winter:1952zz,Cleveland:1975zz,NEMO:1994sst,Arnold:2018tmo,EXO-200:2019rkq,NEXT:2023daz} and
including early cloud chamber and streamer chamber experiments, tracker calorimeters 
and TPCs. Early experimental searches around 1950s-1970s relied primarily on radiochemical,
geochemical  and calorimetric techniques.
Although these pioneering efforts lacked proper event reconstruction and were conducted
without deep underground shielding, they established the experimental feasibility of probing
very rare nuclear transitions.
The subsequent transition to underground laboratory semiconductor, calorimetric  and tracking detectors
from around 1980s delineated a qualitative advance in background suppression, energy resolution 
and overall sensitivity. These form key considerations of modern $0\nu\beta\beta$ searches.

A broad range of nuclear isotopes and
detection technologies have been explored in the context of $0\nu\beta\beta$ searches~\cite{ParticleDataGroup:2024cfk}. For multiple leading isotopes $0\nu\beta\beta$ half-life is restricted to exceed $\mathcal{O}(10^{26})~\mathrm{yr}$.
Within the conventional interpretation based on light Majorana neutrino exchange such bounds
correspond to sensitivity to the effective Majorana mass $|m_{\beta\beta}|$ at the level of
tens of meV, with the precise mapping depending on nuclear matrix element calculations
and phase-space factors~\cite{GERDA:2020xhi,KamLAND-Zen:2022tow,LEGEND:2025jwu}.
Notably, such sensitivity regimes encompass  the parameter space associated with the inverted neutrino
mass ordering and overlap with scenarios with quasi-degenerate neutrino masses.

Extensive historical experimental program searching for $0\nu\beta\beta$ has established
the foundations for increasingly sensitive tests of $L$ violation.
Representative realizations of detection paradigms include high purity germanium
semiconductor detectors such as the LEGEND program~\cite{LEGEND:2025jwu},
isotope-loaded liquid scintillator detectors  such as KamLAND-Zen~\cite{KamLAND-Zen:2024eml}
and SNO+~\cite{SNO:2021xpa},
cryogenic calorimetric experiments such as CUPID~\cite{CUPID:2022jlk}
and AMoRE~\cite{AMoRE:2024loj}, as well as
TPCs with event topology reconstruction, including xenon-based
programs such as nEXO~\cite{nEXO:2021ujk} and gas phase detectors such as NEXT~\cite{NEXT:2023daz}.
These efforts illustrate the continued emphasis on complementarity of experimental
techniques in searches for $\Delta L = 2$ processes.

This highlights that $0\nu\beta\beta$  provides a  link between fundamental neutrino properties, the origin of masses  and possible connections to mechanisms for generating cosmic matter-antimatter asymmetry.
Interpretations of $0\nu\beta\beta$ results benefit  from complementary information on the absolute neutrino mass scale from direct kinematic probes and cosmological observations.
The interpretation of experimental results is limited by uncertainties in nuclear matrix elements, which arise from nuclear many-body correlations, operator renormalization  and short range contributions.
While substantial progress has been achieved through EFT approaches, lattice QCD inputs and ab-initio nuclear structure calculations,  quantitative uncertainties in nuclear matrix elements remain within a factor of few level~\cite{Engel:2016xgb,Cirigliano:2022oqy}.

\subsection{Complementarity of detection techniques}

\begin{table}[t]
\centering
\label{tab:0vbb_complementarity}
\begin{tabular}{C{2.8cm}| P{4cm} P{4cm} P{3.4cm}}
\Xhline{1.2pt}
\multicolumn{1}{C{2.8cm}|}{\makecell{\textbf{Detector}\\\textbf{Class}}} &
\multicolumn{1}{>{\centering\arraybackslash}m{4cm}}{\makecell[c]{\textbf{Primary}\\\textbf{Observable}}} &
\multicolumn{1}{>{\centering\arraybackslash}m{4cm}}{\makecell[c]{\textbf{Background}\\\textbf{Rejection}}} &
\multicolumn{1}{>{\centering\arraybackslash}m{3.4cm}}{\makecell[c]{\textbf{Key}\\\textbf{Features}}} \\

\Xhline{0.8pt}
\multirow{4}{*}{\makecell{High purity\\ semiconductors}}
& Precise summed emitted electron energy 
& Excellent energy resolution, discrimination of spectral backgrounds
& High energy resolution,  well-controlled systematic uncertainties \\
\hline
\multirow{3}{*}{\makecell{Cryogenic\\ calorimeters}}
& Total deposited energy with source embedded in detector
& Narrow energy window tested, low intrinsic radioactive backgrounds
& High resolution, isotope flexibility, multiple targets feasible \\
\hline
\multirow{3}{*}{\makecell{Liquid\\scintillators}}
& Summed electron energy spectrum in large fiducial volumes
& Statistical separation of signal from backgrounds via spectral shape
& Large target masses, long exposures, high detection efficiency \\
\hline
\multirow{3}{*}{\makecell{Tracking and\\ topology detectors}}
& Individual electron tracks, vertices
& Event topology, kinematics, particle identification
& Additional kinematic and topological observables  \\
\hline
\multirow{3}{*}{\makecell{Radiochemical\\ and geochemical}}
& Decay products accumulated over long timescales
& Mode-independent sensitivity to rare processes  
& Integrated sensitivity over long exposure times  \\
\Xhline{1.2pt}
\end{tabular}
\caption{Complementarity of representative experimental approaches for $L$-violating $0\nu\beta\beta$ searches.
All techniques probe  $\Delta L = 2$ nuclear transition but emphasize complementary experimental aspects.}
\label{tab:lviolcomp}
\end{table}

Experimental programs searching for $L$ violation leverage  a diverse range of techniques and detectors, as highlighted by searches for $0\nu\beta\beta$ decay~\cite{Avignone:2007fu,DellOro:2016tmg}. These searches employ a variety of nuclear isotopes, including
$^{76}$Ge, $^{136}$Xe, $^{130}$Te, $^{100}$Mo, and $^{82}$Se,
realized through diverse experimental approaches.
Representative strategies include high purity germanium semiconductor detectors as realized in experiments such as GERDA~\cite{GERDA:2020xhi}, cryogenic calorimeters as employed by CUORE~\cite{CUORE:2024ikf}, liquid scintillator detectors such as KamLAND-Zen~\cite{KamLAND-Zen:2022tow} as well as tracking detectors exemplified by NEMO-3~\cite{Arnold:2018tmo}.
Together these approaches are sensitive to complementary experimental observables and test
distinct aspects of the underlying dynamics related to $L$ violation~\cite{Cirigliano:2022oqy}.

Tab.~\ref{tab:lviolcomp} showcases the complementarity of experimental approaches. This arises from their differences in aspects such as energy resolution, event topologies,
background discrimination techniques, scalability to incorporate large isotope masses as well as
sensitivity to nuclear matrix elements.
Calorimetric detectors allow for excellent energy resolution and provide precise measurements
of the mono-energetic  signatures expected for $0\nu\beta\beta$ decays and hence robust rejection of backgrounds with distributed energy spectra.
Tracking experiments allow reconstruction of event topology and enable powerful
background discrimination with access to kinematic observables of the process, such as emitted electron angular
correlations.
Liquid scintillator detectors can achieve significant exposures and high detection efficiency,
providing strong statistical sensitivity through spectral analyses over long operating experimental periods.
Cryogenic calorimeters combine high energy resolution with low intrinsic backgrounds and
support searches in multiple isotopes within a common experimental framework.

\begin{figure}[t]
\centering
\includegraphics[width=0.65\textwidth]{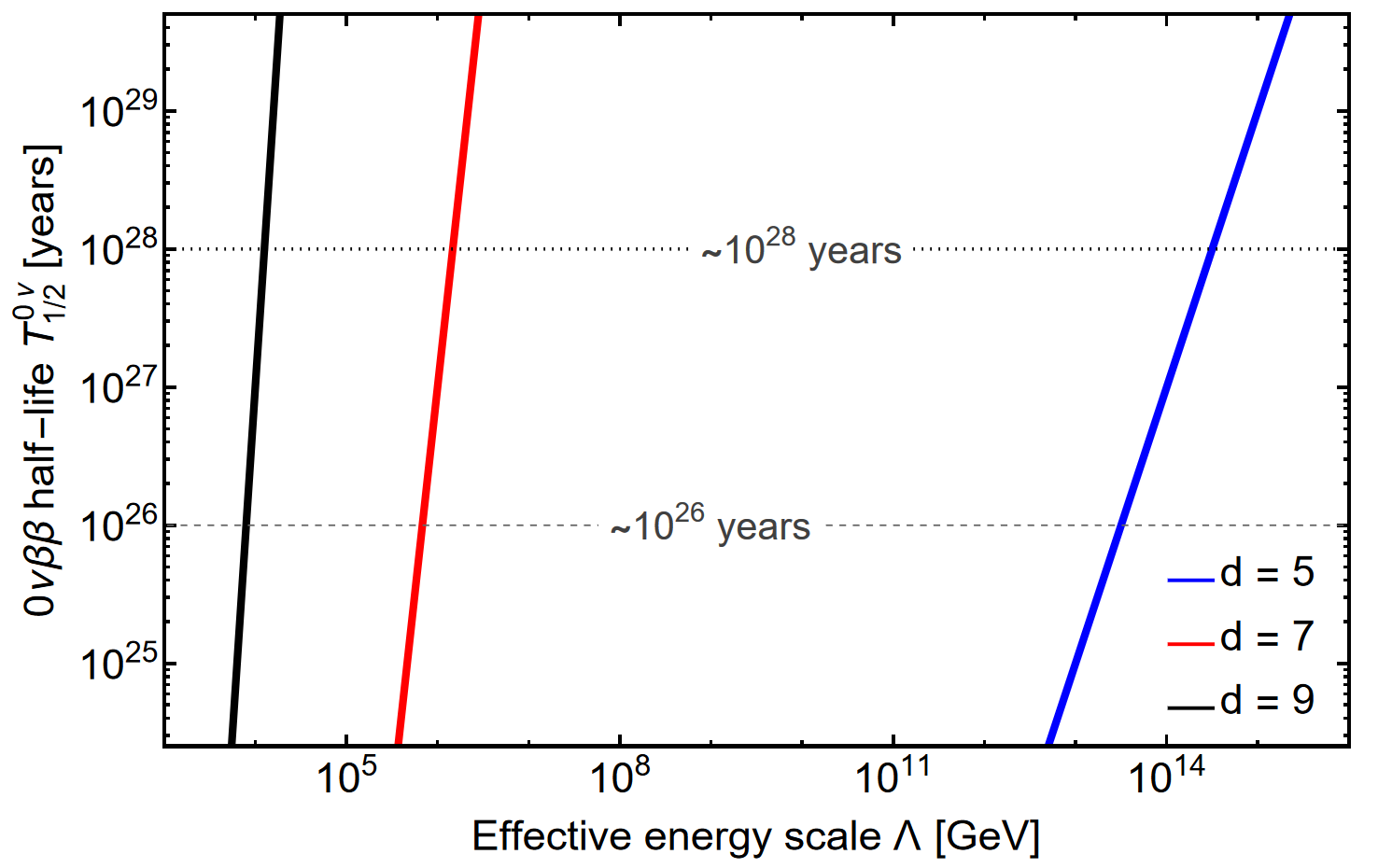}
\caption{Illustrative mapping between the $0\nu\beta\beta$ half-life $T_{1/2}^{0\nu}$ and the effective energy scale $\Lambda$ of $L$-violating physics for different classes of $\Delta L = 2$ operators. The curves represent the parametric EFT scaling with operator dimension $d = 5, 7, 9$. Effects of different nuclear matrix elements, phase-space factors and Wilson coefficients are absorbed into an illustrative normalization. Horizontal lines (dashed, dotted) indicate representative benchmark half-lives. If $L$ violation is observed, the measured half-life maps onto a characteristic effective scale $\Lambda$ for each mechanism and operator class.
}
\label{fig:0vbbscale}
\end{figure}

Fig.~\ref{fig:0vbbscale} illustrates in the context of EFT
how $0\nu\beta\beta$ decay searches provide excellent sensitivity to $L$-violating physics. 
For $\Delta L = 2$ interactions processes contributing to $0\nu\beta\beta$ decay arise from low energy EFT description from operators of different mass dimensions $d$, with decay rates scaling as $\Gamma \propto \Lambda^{-2(d-4)}$ up to contributions from nuclear matrix elements, phase-space factors  and Wilson coefficients as described in Eq.~\eqref{eq:ratescale}. 
Since the leading $\Delta L = 2$ interaction appears at $d = 5$ through the Weinberg operator, experimental
reach in half-life directly translates into sensitivity to   high effective scales that are typically challenging to access.
On the other hand, higher dimensional operators corresponding to short-range mediating mechanisms exhibit a significantly
weaker scaling with half-life $T_{1/2}^{0\nu}$. This motivates the use of complementary isotopes and
detection techniques to disentangle the underlying dynamics in the event of a signal.

Across detection techniques the interpretation of potential experimental signals 
relies on nuclear many-body calculations and phase-space factors.
Using distinct isotopes and detection principles enables mitigation of theoretical uncertainties
and enhances sensitivity to both long-range mechanisms for $0\nu\beta\beta$ decay, such as light Majorana neutrino
exchange, and short-range mechanisms that are associated with higher-dimensional operators.
Consequently, coordinated deployment of complementary detector technologies plays a prominent role in searches for $L$ violation, enabling validation of potential discoveries and discrimination between underlying governing structures and mechanisms.

\subsection{Other $\Delta L \geq 2$ processes and experimental tests}

At low energies $0\nu\beta\beta$ enables sensitive
tests of $L$ violation with $\Delta L = 2$, and the discussion above has focused on
complementary experimental strategies for such searches.
More broadly, however, there exists a broad class of physical processes that can probe
$\Delta L \geq 2$ interactions across complementary kinematic, energy  and experimental regimes.
These include rare meson decays, high energy collider observables and lepton-lepton scattering
processes. Observation of such processes would establish the violation of $L$ and be suggestive of the presence of Majorana fermions or other sources of
$L$-violating interactions beyond SM.
Notably, these processes probe combinations of couplings, flavor structures  and energy
scales that are distinct from, and complementary to, those accessed by $0\nu\beta\beta$.

The essence of this complementarity can be understood by noting that different
$L$-violating processes probe distinct characteristic momentum transfers $q$.
This is related to the relevant physical length scales of the processes.
In particular, $0\nu\beta\beta$ involves typical virtual momenta of
$q \sim \mathcal{O}(100~\mathrm{MeV})$ that are characteristic of distances inside nuclei~\cite{Cirigliano:2022oqy}.
Consequently, this is sensitive both to long range mechanisms mediated by light Majorana neutrinos
of mass $M$, with amplitudes scaling schematically as $\sim 1/(q^2 + M^2)$, and to short range contact-type
interactions induced by heavy new physics, scaling as $\sim 1/M^2$
after integrating out heavy degrees of freedom
\cite{Schechter:1981cv,Engel:2016xgb}.
On the other hand, rare meson decays and lepton-lepton scatterings can probe momentum transfers in the
range $q \sim \mathcal{O}(\mathrm{MeV}-\mathrm{GeV})$ and can access different
$\Delta L \geq 2$ effective operators with distinct Lorentz and flavor structures
\cite{deGouvea:2013zba}.
Energetic collider searches further extend sensitivity to momentum transfers at and above the
electroweak scale $\mathcal{O}(\mathrm{TeV})$,  where heavy Majorana fermions or mediators of other
$L$-violating interactions may be produced directly and enable probing dynamics at shorter distances~\cite{Kersten:2007vk,Atre:2009rg}.
As in the case of $B$ violation,  the EFT  framework provides a
unified description for relating these regimes and for systematically identifying potential correlations among different experimental probes.

Searches for rare processes using intense beams can probe same sign di-lepton $l$ decays of charged
mesons, such as
$K^+ \to \pi^- \ell^+ \ell^+$,
$D^+ \to \pi^- \ell^+ \ell^+$, and
$B^+ \to D^- \ell^+ \ell^+$,
where $\ell = e,\mu$.
These $\Delta L = 2$ transitions can be mediated by  Majorana right handed neutrinos (sterile neutrinos) or,
more generally, by  $L$-violating operators involving both quark and
lepton fields.
Dedicated searches including at kaon, charm, and $B$-meson experiments, such as NA62
\cite{NA62:2021bji}, LHCb~\cite{LHCb:2014osd}  and Belle~II \cite{Belle-II:2018jsg}, place stringent limits on
their branching fractions that are typically at or below the level
${\rm Br} \lesssim \mathcal{O}(10^{-8})$.
These results constrain combinations of heavy right handed neutrino masses and active-sterile
neutrino mixing parameters in the MeV-GeV mass range~\cite{Atre:2009rg}.
Related low energy probes include inverse beta-type nuclear transitions
($nn \to pp e^- e^-$) that test $\Delta L = 2$ quark-lepton interactions at nuclear distance scales, and muonium-antimuonium oscillations
($\mu^+ e^- \to \mu^- e^+$) that probe related $L$-violating four-lepton interactions at atomic distance scales 
\cite{ParticleDataGroup:2024cfk,Willmann:1998gd}.
Such searches can be especially sensitive to scenarios with intermediate mass right handed neutrinos
that may be challenging to access with both $0\nu\beta\beta$ decays and high energy collider experiments.

At higher energies  hadron collider experiments can probe same sign di-lepton events such as $  
pp \to \ell^\pm \ell^\pm jj$, where $jj$ denotes jets. These processes 
can arise when a heavy Majorana right handed neutrino $N$ is produced and subsequently decays via
$N \to \ell W$, with $W$ being the SM $W$-boson.
As exemplified by analyses at ATLAS \cite{ATLAS:2022atq} and CMS \cite{CMS:2021dzb}, hadronic collisions enable probing $\Delta L = 2$
interactions at the electroweak and TeV scales and can be sensitive to the flavor structure of
heavy neutrino couplings.
Sizable active-sterile
mixings for heavy neutrino masses up to the electroweak scale are constrained, while  higher energy or
higher luminosity facilities could extend sensitivities to heavier mediators or smaller couplings. More generally, collider probes can also be sensitive to
other sources of $L$ violation, including scalar or vector mediators and distinct
EFT operators.
In some regimes collider constraints can be
competitive with or stronger than those derived from nuclear decays, although for the
contact-type interactions $0\nu\beta\beta$ decays are particularly sensitive.
Processes such as electron-electron scattering 
$e^- e^- \to W^- W^-$ 
can offer a clean $\Delta L = 2$ probe at luminous lepton colliders~\cite{London:1987nz}, as $L$-violating interactions are tested without hadronic uncertainties.
Such observations could provide a more direct test of the possible
Majorana nature of neutrinos and probe momentum transfers far above those accessible in nuclear
systems.

Taken together, rare meson decays, collider searches and lepton-lepton scattering provide
powerful and complementary probes of $\Delta L \geq 2$ interactions.
The non-observation of such processes   disfavors low scale or unsuppressed sources
of $L$ violation. On the other hand, discovery would have profound implications for the origin
of neutrino masses and the organizing symmetries of fundamental interactions.
Such probes play an important role in interpreting potential observations of
neutrinoless $0\nu\beta\beta$ and in discriminating among underlying microscopic mechanisms
of $L$ violation.

 \subsection{Cosmological and astrophysical constraints and probes}

Cosmology provides powerful and complementary settings for constraining neutrino properties and in model-dependent settings scenarios associated with $L$ violation through the collective impact of neutrinos on the evolution of the Universe.  
Although cosmological observables do not typically resolve individual microscopic $L$-violating processes, they are sensitive to the absolute neutrino masses, cosmic neutrino background (C$\nu$B) and the possible presence of a primordial lepton asymmetry.  
These quantities are connected to the underlying physics that also motivate laboratory searches for $L$ violation, complementing each other.

A primordial lepton asymmetry can be generated in the early Universe through various scenarios, such as decays of heavy right handed neutrinos~\cite{Fukugita:1986hr} or Affleck-Dine mechanism~\cite{Affleck:1984fy,Dine:1995kz}. It affects the energy density and weak interaction rates in the early Universe.  
A lepton asymmetry in the neutrino sector contributes to the effective number of relativistic degrees of freedom~ $N_{\rm eff}$.  
In the SM, including non-instantaneous weak decoupling and finite temperature effects, it is found that $N_{\rm eff} \simeq 3.045$~\cite{Mangano:2005cc}. Non-negligible lepton asymmetry  can shift this value and modify both Big Bang nucleosynthesis (BBN) predictions for light element abundances and the cosmic microwave background (CMB) anisotropy spectrum, thereby providing sensitive constraints on deviations from standard cosmological evolution~\cite{Iocco:2008va,Froustey:2024mgf}.  
Measurements of primordial element abundances and precision CMB observations stringently constrain such deviations, implying that any surviving lepton asymmetry at the epoch of BBN is tightly constrained.

The potential existence of a primordial lepton asymmetry can play a central role in a broad class of scenarios for generating the observed cosmological baryon asymmetry through leptogenesis~\cite{Fukugita:1986hr}. Here, $L$-violating dynamics in the early Universe generate an initial asymmetry that is later partially converted into the observed baryon asymmetry through electroweak sphaleron processes.  
Further, lepton asymmetry can play a role in certain model-dependent scenarios for generating dark matter abundance,  such as resonant active-sterile neutrino oscillations~\cite{Shi:1998km}.

Analyses of cosmological datasets provide some of the most stringent constraints on the absolute neutrino mass scale and related scenarios associated with $L$ violation, although with additional underlying assumptions compared to direct laboratory searches and are therefore complementary.
Once neutrinos become non-relativistic  their finite masses suppress the growth of matter perturbations and structure on small scales, leaving characteristic imprints in larger scale structure, gravitational lensing  and the CMB power spectrum.  
Cosmological analyses can restrict the sum of neutrino masses typically to levels of $\sum m_\nu \lesssim \mathcal{O}(0.1) \mathrm{eV}$~\cite{Lesgourgues:2012uu,Planck:2018vyg,ParticleDataGroup:2024cfk}, depending on the choice of datasets and modeling assumptions. 
These tests are complementary to direct kinematic measurements of neutrino masses and to searches for $0\nu\beta\beta$, which probe the Majorana nature of neutrinos and hence explicit $\Delta L = 2$ violation.

Astrophysical settings, such as core-collapse supernovae, provide complementary sensitivity to exotic neutrino properties that may be associated with $L$ violation through the dynamics of neutrinos and their propagation~\cite{Raffelt:1996wa}. In such extreme stellar environments neutrinos carry away vast amounts of energy and exotic neutrino properties could in principle affect emission, cooling and other aspects. This is illustrated with detected neutrino emission from SN 1987a~\cite{Kamiokande-II:1987idp,Bionta:1987qt}, which constrains some nonstandard neutrino properties.  Broader astronomical constraints, including searches for signals from potential neutrino decays and from diffuse astrophysical neutrino backgrounds~\cite{DeGouvea:2020ang}, further delineate viable theoretical parameter spaces. 

Cosmological observations thus provide a broad and powerful set of indirect probes of $L$ violation and neutrino properties.  
They complement laboratory searches, such as  neutrino oscillation experiments, CLFV and $0\nu\beta\beta$ decays, by probing the collective impact of neutrinos on the evolution  of the Universe and its structure. Astrophysical tests are also synergistic by probing $L$ violation in  localized and extreme environments. Together, these complementary searches constrain the landscape of viable $L$-violating scenarios and highlight the role of such processes in shaping fundamental physics.

\section{Interpreting Tests of Baryon and Lepton Number Conservation} 

\subsection{Implications of discovery and non-discovery}

Experimental tests of $B$ and $L$ conservation probe whether the apparent stability of ordinary matter reflects an exact symmetry structure of Nature or instead an emergent low energy consequence of SM field content and gauge symmetries. These searches can be sensitive to extreme effective energy scales. Consequently, both discovery and non-discovery can have direct implications for the underlying structure of microphysics at higher energies, as well as fundamental questions such as the origin of neutrino masses and the viability of broad classes of cosmological early Universe scenarios.

A confirmed observation of $B$ violation in nucleon or nuclear processes would establish that $B$ is not an exact symmetry protecting ordinary   matter. In empirical terms this would demonstrate that baryonic matter is not absolutely stable, with a finite lifetime governed by effective interactions beyond the renormalizable interactions of SM. This alone would constitute a qualitative and significant discovery and reveal the existence of $\Delta B\neq0$ interactions. However, what specifically would such discovery reveal about microphysics and underlying structure depends strongly on observed physical processes and their details, including content of the final states, kinematics and process branching patterns. In the context of an EFT interpretation these features restrict the relevant quantum numbers and dimensionality of the leading contributing operators, thus suggesting the characteristic effective mass scale and associated selection rules that include information about whether $\Delta(B-L)=0$ or $\Delta(B-L)\neq0$ is preferred~\cite{Weinberg:1979sa,Wilczek:1979hc}. Certain 
patterns of physical processes can be suggestive of underlying quark-lepton unification structures, while others may point to alternative organizing principles such as additional symmetries, flavor structure or additional dynamics. Notably, nucleon decay by itself would not constitute a unique confirmation of a specific underlying GUT. The most robust output of a $B$ violation discovery would be establishment that there are $\Delta B\neq0$ interactions, constraints on EFT operator classes obtained from information about kinematics and physical process  as well as infer an effective energy scale that can be compared with the expectations from unification, flavor structures and potential underlying microphysics scenarios~\cite{Georgi:1974sy,Pati:1973uk}. 

The continued non-observation of $B$ violation provides useful information and has played a central role in shaping theoretical efforts to uncover deeper organizing principles of fundamental interactions. This is well illustrated by constraints on nucleon decay, particularly in channels such as $p\to e^+\pi^0$ that provided
decisive experimental falsification of the minimal $SU(5)$ GUT~\cite{Georgi:1974sy} under conventional assumptions. Early results at large underground detectors, notably Kamiokande~\cite{Kamiokande-II:1989avz} and IMB~\cite{McGrew:1999nd}, established that the proton lifetimes predicted in such minimal realizations~\cite{Ellis:1980jm} were incompatible with observations. This was reinforced with subsequent observations. Minimal  $SU(5)$ models with TeV-scale SUSY are also strongly constrained by proton decay searches, in particular by limits on channels such as $p \to \bar{\nu} K^+$ from Super-Kamiokande~\cite{Super-Kamiokande:1999xld}. In their simplest realizations  and under the assumptions specified in Ref.~\cite{Murayama:2001ur} such models are excluded or strongly disfavored. Compatibility with experimental data in these frameworks requires additional suppression mechanisms and more complex structures, which may include symmetries such as SUSY realized at different mass scales, non-trivial flavor structures or non-minimal field contents. 
Beyond simplest two-body nucleon decay channels, multi-body and multi-nucleon decays play an important complementary role. This is exemplified by  $p \rightarrow e^+\nu\nu$ with $|\Delta(B-L) = 2|$, where Super-Kamiokande
searches~\cite{Super-Kamiokande:2014pqx} have further constrained classes of $B$-violating mechanisms in model-dependent ways, including scenarios with extended scalar sectors or non-minimal symmetry breaking patterns~\cite{Gu:2011pf,Pati:1983jk}.
This illustrates how null results enable systematic elimination of minimal theoretical models and restrict distinct $B$ violation mechanisms, favoring realizations with higher effective mass scales or additional protective structures that suppress observable $B$-violating processes.

A confirmed observation of $L$ violation, most notably through $0\nu\beta\beta$ decay, would establish that $\Delta L=2$ at low energies and imply that $L$ is not an exact symmetry. This would imply that a Majorana mass term for neutrinos is generated in the particular sense guaranteed by the Schechter-Valle theorem~\cite{Schechter:1981bd}. However, this observation by itself will not uniquely determine the microscopic mechanism responsible for the physical process. The interpretation of a measured rate is intrinsically dependent on the considered mechanism. In particular, for $0\nu\beta\beta$ decay long range light neutrino exchange corresponds to an effective Majorana mass parameter~\cite{Bilenky:1987ty}, whereas short range $\Delta L=2$ interactions can arise from higher dimensional operators with different scaling behavior and possibly different correlations with other observables~\cite{Prezeau:2003xn,Deppisch:2012nb}. A discovery would therefore primarily establish the existence of $\Delta L=2$ interactions and provide quantitative information on the effective energy scale and structure of the responsible operators when appropriately combined with nuclear and kinematic structure. Further, this would demonstrate that $L$ violation is a potential ingredient for early Universe dynamics. While $\Delta L=2$ is a common ingredient for typical leptogenesis frameworks for generating the cosmic baryon asymmetry~\cite{Fukugita:1986hr}, observation of $0\nu\beta\beta$ decay alone would not establish that leptogenesis occurred. Instead, this could empower classes of baryogenesis scenarios in which $L$ violation plays a central role, but leave alternative mechanisms open~\cite{Sakharov:1967dj}.

As in the case of $B$ violation, continued non-observation of $L$ violation provides valuable information and  guidance for theoretical efforts aiming to reveal potential deeper underlying principles.   
Null results provide stringent constraints for minimal scenarios in which TeV-scale $L$-violating dynamics dominate $0\nu\beta\beta$ decay with unsuppressed couplings and disfavor minimal low energy scale realizations without additional suppression, such as from additional structure.
Further, null results from $0\nu\beta\beta$ decays also restrict interpretations based on light sterile neutrino exchange with sizable mixings. 
Notably, increasingly stringent limits from $0\nu\beta\beta$ decay searches disfavor some minimal low energy realizations of $L$ violation with unsuppressed couplings and can constrain specific regions of parameter space in Majorana neutrino mass scenarios, while remaining compatible with high scale or otherwise suppressed origins of neutrino mass and $L$ violation including high scale leptogenesis frameworks~\cite{Davidson:2008bu}. In this context, continuous non-observation of $0\nu\beta\beta$ decay can favor either high energy scale origins of $L$ violation, suppressed operator coefficients  or structured cancellations among possible contributions to the process.

Given the extreme rarity of the processes involved any claimed observation of $B$ violation or $L$ violation necessitates independent confirmation for a rigorous interpretation. The history of rare event searches includes instances in which early indications in both nucleon decay and $0\nu\beta\beta$ decay were not corroborated by subsequent experimental observations. This underscores the significance of replication, complementary detection techniques  and consistency across experimental targets and physical processes under investigation. More broadly, the enduring implications of searches for $B$ violation and $L$ violation are not restricted to any single experimental milestone, but to the role these symmetries play in fundamental theory.
These searches exhibit deep complementarity with other probes of fundamental physics, including high energy collider experiments, precision measurements with intense sources and sensitive detectors as well as astrophysical and cosmological observations. A consistent pattern of signals or constraints across such diverse probes could greatly enhance interpretations and assist in discriminating among competing potential underlying structures. Discovery of described processes would establish the existence of new interactions violating these symmetries and restrict their operator structure and effective characteristic energy scale. On the other hand, continuous observations with non-discovery probes higher energy scales at which such interactions may appear and could favor underlying frameworks where   violations are more suppressed.
In both cases, these experimental searches provide deep insights into fundamental physics including
the stability of matter, the nature of neutrino mass  and the connection between laboratory observables and the high energy organizing principles of particle physics and cosmology.

 \subsection{Map for discovery interpretation}

\begin{figure}[t]
\centering
\begin{tikzpicture}[
  box/.style={draw, rounded corners, align=left, inner sep=6pt, text width=4.5cm},
  topbox/.style={draw, rounded corners, align=center, inner sep=6pt, text width=10cm},
  midbox/.style={draw, rounded corners, align=left, inner sep=6pt, text width=4.5cm},
  decision/.style={draw, rounded corners, align=center, inner sep=6pt, text width=4.5cm},
  arrow/.style={-Latex, line width=0.6pt},
  node distance=6mm
]
\node[topbox] (obs) {\textbf{Observation of Global Charge Violation}\\
(evidence for non-conservation of baryon $(B)$, lepton $(L)$ number)};
\node[topbox, below=6mm of obs] (q) {\textbf{Which symmetry violation is experimentally established?}};
\matrix (M) [matrix of nodes,
  nodes in empty cells,
  column sep=7mm,
  row sep=6mm,
  below=4mm of q
]{\node[box] (b) {\centering\textbf{$\Delta B \neq 0$ observed}\\
  (e.g.  nucleon decay, $n - \bar{n}$, multinucleon processes)}; &
  &
  \node[box] (l) {\centering\textbf{$\Delta L \neq 0$ observed}\\
  (e.g.  $\Delta L=2$ via $0\nu\beta\beta$ transitions, other  processes)}; \\ 
  \node[decision] (b_dec) {Is there also independent evidence for $\Delta L \neq 0$? };   & 
  \node[midbox, yshift=-10mm] (both) {\centering\textbf{Both $\Delta B \neq 0$ and $\Delta L \neq 0$ observed}\\
Tests if $B - L$ is conserved or violated. Constrains the symmetry structure and determines allowed effective operator classes.
Can possibly have implications for viable mechanisms for the origin of the cosmic baryon asymmetry.}; &
  \node[decision] (l_dec) {Is there also independent
evidence for $\Delta B \neq 0$?}; \\
};
\node[box, below=10mm of b_dec, yshift=-10mm] (b_only) {\centering\textbf{$\Delta B \neq 0$ only}\\ \raggedright
Implies baryonic matter is not absolutely stable. Determines allowed effective operator classes, $\Delta(B-L)$ may be zero or not depending on dominant process. Constrains the characteristic $B$-violation mass scale of the underlying effective operators.};
\node[box, below=10mm of l_dec, yshift=-10mm] (l_only) {\centering\textbf{$\Delta L \neq 0$ only}\\
Implies $L$ is not exact. Determines allowed effective
operator classes, for  $\Delta L=2$ one has $\Delta(B-L)\neq 0$ and suggests Majorana neutrino masses. Constrains the characteristic $L$-violation mass scale of the underlying effective operators.};
\draw[arrow] (obs) -- (q);
\coordinate (qL) at ([xshift=-2.4cm,yshift=-1mm] q.south);
\coordinate (qR) at ([xshift= 2.4cm,yshift=-1mm] q.south);
\draw[arrow] (q) -- (b.east);
\draw[arrow] (q) -- (l.west);
\draw[arrow] (b) -- (b_dec);
\draw[arrow] (l) -- (l_dec);
\draw[arrow] (b_dec) -- node[midway, left] {\small NO} (b_only);
\draw[arrow] (l_dec) -- node[midway, right] {\small NO} (l_only);
\draw[arrow] (b_dec.east) -- (both.west);
\node[anchor=west] at ($(b_dec.east)!0.55!(both.west) + (-10mm,0mm)$) {\small YES};
\draw[arrow] (l_dec.west) -- (both.east);
\node[anchor=east] at ($(l_dec.west)!0.55!(both.east) + (10mm,0mm)$) {\small YES};
\end{tikzpicture}
\caption{Discovery decision map for implications of observations of $B$ violation and $L$ violation.}
\label{fig:discoverymap}
\end{figure}
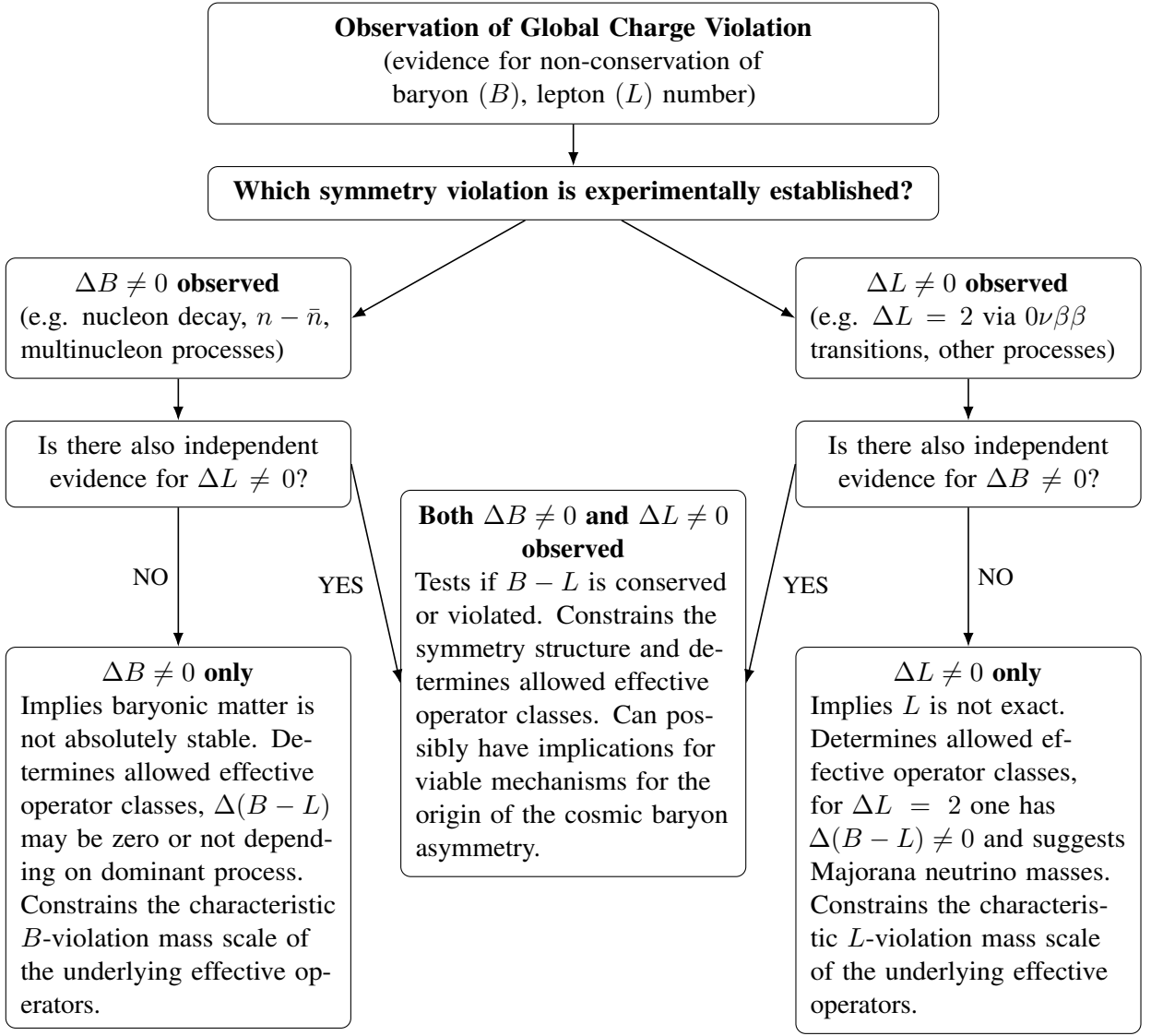

While the preceding discussion has focused on the implications of discovery or non-discovery of $B$ violation and $L$ violation in generality, the interpretation of potential experimental signals can be organized into a concise logical decision framework. The observation of $B$-violating or $L$-violating processes would constitute a fundamental departure from the accidental global symmetries of SM. Interpreting such a discovery therefore requires identifying which global charges are violated and how the observed processes relate to the potential underlying structures. The pattern of observed $B$ violation and $L$ violation provides robust  information based on symmetries that can be assessed independently of specific high energy microphysics theory realizations.

Fig.~\ref{fig:discoverymap} presents a schematic discovery decision map that summarizes how different combinations of experimentally established global charge violations guide the interpretations of possible underlying physics. The map separates cases in which $\Delta B\neq0$, $\Delta L\neq0$  or both are observed. Further, it illustrates how the combined pattern of $B$-violating and $L$-violating signatures  or their absence allows discriminating among broad theoretical frameworks. Observation of $\Delta B=1$ processes without evidence for $\Delta L=2$ would point toward scenarios in which $B$ violation and $L$ violation are not tightly linked at experimentally accessible scales. Conversely, evidence for $\Delta L=2$ without $B$ violation would indicate that $L$ is violated in parts of underlying theory or at energy scales that do not typically induce observable nucleon decays. Joint evidence for both $B$ violation and $L$ violation would motivate potential unified descriptions or correlated symmetry breaking patterns, but the mapping from low energy observables to high energy underlying theory would remain non-unique and would rely on consistency across physical processes, experimental targets  and complementary probes.

In case $B$ violation is observed in the absence of any evidence for $L$ violation, the data would be consistent with interactions that conserve the combination $B-L$ in many frameworks. In many classes of theoretical model realizations the responsible operators respect $B-L$ symmetry and are associated with high characteristic mass scales. Conversely, the observation of $L$ violation, such as $\Delta L=2$ transitions with $0\nu\beta\beta$ decay, would directly imply non-conservation of $B-L$ and point toward qualitatively different underlying symmetry structures. This carries significant implications for the origin of neutrino masses. The simultaneous observation of both $B$-violating and $L$-violating processes would further improve the interpretation by constraining the possible allowed operators and symmetry structures of the underlying theory, as well as by providing insights into whether $B$ violation and $L$ violation are related to a common dynamical origin or arise from distinct mechanisms.

Notably, the discovery map is not intended to identify a unique underlying theoretical model. Rather, it encodes conclusions based on symmetries that follow directly from the pattern of observed global charge violation, without additional assumptions regarding possible dominant operator dimension, process branching fractions  or  mediator information. Consequently, the map remains  applicable   regardless of the specific process and channel in which a discovery occurs and provides a durable guide for interpreting future observations within the broader landscape of physics beyond SM.

\section{Summary}

Tests of $B$ and $L$ conservation occupy a distinct position in the landscape of fundamental physics. Unlike a multitude of searches that target specific particles or energy scales these investigations probe the underlying organizing principles themselves, addressing whether the apparent stability of matter and the structure of the leptonic interactions reflect exact symmetries or emergent low energy patterns. $B$ violation and $L$ violation can naturally appear across a wide range of theoretical frameworks and are closely connected to fundamental topics including unification, the origin of neutrino mass  and cosmic matter-antimatter asymmetry.

From experimental perspective  searches for $B$ violation and $L$ violation highlight the power of rare event techniques. Combining large target masses, long experimental exposure times and advanced background rejection such searches enable accessing effective energy scales far beyond typical laboratory reach. Distinct processes, such as nucleon decay, $n-\bar{n}$ transformations and $0\nu\beta\beta$ decay, allow probing complementary interaction structures and governing symmetry patterns. Notably, no single physical process and channel or technology is by itself entirely sufficient for comprehensive exploration of $B$ violation and $L$ violation. A diversified experimental program is highly advantageous. This allows to maximize discovery potential and enables robust interpretations of observations.

The discovery of $B$ violation or $L$ violation would represent a qualitative advance in fundamental physics. Observation of $B$ violation would demonstrate that the stability of ordinary baryonic matter is not protected by an exact symmetry. Observation of $L$ violation, particularly by two units, would imply the generation of a Majorana mass contribution on general quantum field theory grounds and provide structural insights into the origin of mass in the neutral leptons. Experimental signal would not be interpreted in isolation. The interpretation would rely on correlations across multiple physical processes, energy scales  and experimental platforms, as well as linking laboratory observations to cosmology and fundamental theory.

The absence of observed violation  is itself highly informative. Increasingly stringent limits constrain broad classes of theories, favoring viable scenarios associated with higher effective energy scales, more intricate symmetry structures and  suppressed interactions. Non-observation thus plays an active and important role in shaping and guiding theoretical understanding, narrowing the space of consistent extensions of known physics and honing the questions posed to future experimental searches.

Looking forward, tests of $B$ and $L$ conservation will continue to evolve with advances in detector technology, nuclear and hadronic theory as well as precision modeling. Their enduring significance is highlighted not only by the possibility of a discovery, but also by their capacity to interrogate deep structural assumptions and structures of fundamental physics.  
Whether through discovery or increasingly precise null results  these searches remain indispensable for understanding how the laws governing matter can emerge, persist  and potentially fail at the most fundamental level.

\section*{Acknowledgments}
The author thanks the editors of the Encyclopedia and numerous colleagues for earlier insightful discussions. The author acknowledges support by the World Premier International Research Center Initiative (WPI), MEXT, Japan.

\bibliographystyle{utmod}
\bibliography{refs}
\end{document}